\documentclass[twocolumn]{aastex631}
\usepackage[utf8]{inputenc}
\usepackage{amsmath}
\usepackage{natbib}
\usepackage{bm}
\usepackage{multirow}
\usepackage{comment}

\newcommand{\p}{\partial}

\newcommand{\ii}{\mathrm{i}}
\newcommand{\C}{\mathcal{C}^\lambda}

\newcommand{\dd}{\delta}

\newcommand{\reynolds}{\operatorname{Re}}
\newcommand{\peclet}{\operatorname{Pe}}
\newcommand{\nrsq}{\mathcal{N}}

\newcommand{\Hdust}{H_\mathrm{d}}
\newcommand{\Hgas}{H_\mathrm{g}}

\newcommand{\hgas}{h_\mathrm{g}}

\newcommand{\rhog}{\rho_\mathrm{g}}

\newcommand{\vg}{\bm{v}_\mathrm{g}}
\newcommand{\vdx}{v_{\mathrm{d}x}}
\newcommand{\vdy}{v_{\mathrm{d}y}}

\newcommand{\vgx}{v_{\mathrm{g}x}}
\newcommand{\vgy}{v_{\mathrm{g}y}}
\newcommand{\vgz}{v_{\mathrm{g}z}}
\newcommand{\Dvx}{\Delta v_x}
\newcommand{\Dvy}{\Delta v_y}
\newcommand{\Dvz}{\Delta v_z}

\newcommand{\taus}{\tau_\mathrm{s}}
\newcommand{\st}{\mathrm{St}}
\newcommand{\w}{ \widetilde{W}}
\newcommand{\fdust}{f_\mathrm{d}}
\newcommand{\fgas}{f_\mathrm{g}}

\newcommand{\tcool}{t_\mathrm{cool}}

\newcommand{\bmv}{\bm{v}}

\newcommand{\re}{\operatorname{Re}}

\newcommand{\rhogref}{\rho_\mathrm{g}}

\newcommand{\Dv}{\Delta\bm{v}}
\newcommand{\vgtilde}{\widetilde{\bm{v}}_\mathrm{g}}
\newcommand{\vdtilde}{\widetilde{\bm{v}}_\mathrm{d}}
\newcommand{\vgtildex}{\widetilde{v}_{\mathrm{g}x}}

\newcommand{\xhat}{\hat{\bm{x}}}
\newcommand{\yhat}{\hat{\bm{y}}}

\newcommand{\ikx}{\mathrm{i}k_x}

\newcommand{\ik}{\mathrm{i}\bm{k}}
\newcommand{\nud}{\nu_\mathrm{d}} 

\newcommand{\Porb}{P_\mathrm{orb}}
\newcommand{\epsinit}{\epsilon_0}
\newcommand{\Fd}{F_\mathrm{d}}
\newcommand{\Fg}{F_\mathrm{g}}

\newcommand{\Pg}{P_\mathrm{g}}

\defcitealias{teed21}{TL21}

\shorttitle{Dust dynamics in non-isothermal gas}
\shortauthors{M.-K. Lin, M. Lehmann}

\begin{document}

\title{Dust dynamics in radially convective regions of protoplanetary disks}

\author[0000-0002-8597-4386]{Min-Kai Lin}
\email{mklin@asiaa.sinica.edu.tw}
\affiliation{Institute of Astronomy and Astrophysics, Academia Sinica, Taipei 10617, Taiwan}
\affiliation{Physics Division, National Center for Theoretical Sciences, Taipei 10617, Taiwan}

\author[0000-0002-0496-3539]{Marius Lehmann}
\affiliation{Institute of Astronomy and Astrophysics, Academia Sinica, Taipei 10617, Taiwan}

\begin{abstract}
Hydrodynamic instabilities likely operate in protoplanetary disks. One candidate, Convective Overstability (COS), can be triggered in regions with a negative radial entropy gradient. The ensuing turbulence and flow structures are expected to affect dust dynamics directly. We revisit the interaction between dust and the COS with high-resolution spectral simulations in the unstratified, axisymmetric Boussinesq shearing box framework. We find zonal flows, or pressure bumps, formed by the COS trap dust, as expected, but {
dust densities increase at most by a factor of $O(10)$ over its background value due to the zonal flows' unsteady nature.}
Furthermore, dust feedback can impede the formation of zonal flows, even at small dust-to-gas ratios $\epsilon \sim O(0.1)$. We interpret this phenomenon as a competition between the negative gas angular momentum flux associated with zonal flow formation and the positive dust angular momentum flux associated with its drift towards pressure maxima. Dust concentration significantly weakens when a large-scale radial pressure gradient induces a background dust drift. Ultimately, we find that dust concentration by COS-induced zonal flows is limited to $\epsilon \lesssim 1$. Whether this can be improved under more realistic geometries must be addressed with stratified and full 3D simulations at equivalent resolutions.  
\end{abstract}
\section{Introduction}\label{intro}

Protoplanetary disks (PPDs) --- the birth sites of planets --- comprise mainly of gas with a sub-dominant solid component by mass \citep{miotello23}. While solids provide the raw material for planet formation \citep{drazkowska23}, they are immersed in a gaseous environment. The gas dynamics of PPDs thus profoundly impact dust evolution. For example, turbulence can stir dust grains, which works against their collisional growth. At the same time, coherent structures such as pressure bumps or vortices can concentrate dust and facilitate their gravitational collapse into planetesimals \citep{birnstiel24}. 

PPDs are prone to several gas dynamical instabilities \citep{fromang17,lyra19,lesur23}. Due to their low ionization levels, magneto-hydrodynamic instabilities are limited in the disk bulk. This has renewed interest in purely hydrodynamical mechanisms for driving turbulence and structure formation. Over the last decade, at least three hydrodynamic instabilities have been discussed: the Vertical Shear Instability \citep[VSI,][]{nelson13,barker15}; the Convective Overstability \citep[COS,][]{klahr14,lyra14}; and the Zombie Vortex Instability \citep[ZVI,][]{marcus13,marcus15}.

The VSI, COS, and ZVI are also termed thermo-hydrodynamic instabilities because of their respective requirements on the gas' thermal or cooling timescales. The VSI requires rapid cooling and applies to the outer disk \citep{lin15}, while the ZVI requires slow cooling and is thus applicable to the innermost disk \citep[][but see \citet{barranco2018}]{lesur16}. On the other hand, intermediate cooling timescales optimize the COS and may apply to planet-forming regions. The COS also requires an entropy profile that decreases outward, which is atypical in { the midplane of radially smooth disks,} but could occur around particular locations such as gap edges or dead zone boundaries. { Alternatively, regions away from the midplane may exhibit a negative radial entropy gradient and develop COS \citep{lesur23}}.  


{ The COS arises from destabilized inertial waves, which are usually restored by the Coriolis force \citep{balbus03} and thus only exist in rotating flows. They also have smaller frequencies than the rotation frequency. However, in COS, a radially oscillating fluid parcel can 
increase its oscillation amplitude if it loses (gains) sufficient heat to its surroundings when it moves outward (inward) \citep{latter16}.} This translates into the above thermal and structural requirements. Under axisymmetry, the COS leads to the formation of zonal flows or pressure bumps \citep[][hereafter \citetalias{teed21}]{teed21}; while in 3D, vortices also form \citep{lyra14,raettig21,lehmann24}. Interestingly, a non-linear, non-axisymmetric version of a similar process, the Sub-critical Baroclinic Instability (SBI), was discovered before the linear COS that can amplify pre-existing vortices \citep{peterson07a,peterson07b,lesur10,lyra11,raettig13}.

\citet{raettig15, raettig21} found that  COS-produced and SBI-sustained vortices can concentrate dust to sufficiently high densities for gravitational collapse into planetesimals \citep{lyra24}. In these studies, gas is modeled in an augmented compressible shearing box \citep{lyra11}. This contrasts with the Boussinesq framework typically used to model convection and can be self-consistently derived \citep{latter17b}. 

In this work, we revisit the problem of dust interacting with the COS in a Boussinesq shearing box. We append dust as a second, pressureless fluid coupled to the gas via drag forces. This approach was also taken in our recent linear analyses of non-isothermal gas-dust interaction \citep{lehmann23}. We showed that dust-loading hampers the COS by reducing the disk's effective buoyancy. Furthermore, finite drag forces set a minimum length scale for instability.

Here, we perform spectral simulations to examine the nonlinear evolution of a dusty gas subject to the COS. Our simulations are dusty extensions to the pure gas COS simulations in \citetalias{teed21}. We find that dust concentration by the COS is dynamic and can be limited by feedback even at low dust-to-gas ratios. We also find that a background dust drift reduces the efficacy of zonal flows to act as dust traps. This suggests it may be difficult to trigger the streaming instability \citep[SI,][]{youdin05,youdin07}  --- the de facto route to planetesimal formation --- by COS-induced zonal flows. 

This paper is organized as follows. We describe the physical setup and basic equations for modeling dusty COS in a Boussinesq shearing box in \S\ref{model}. We briefly review the linear stability of the system in \S\ref{linear}. We describe the numerical approach to simulate its nonlinear evolution in \S\ref{numerical}. We present results in \S\ref{results} and conduct a parameter survey in \S\ref{survey}. We discuss our results in \S\ref{discussion} and conclude in \S\ref{summary}.

\section{Basic equations}\label{model}
We consider a three-dimensional, dusty protoplanetary disk (PPD) orbiting a star of mass $M_*$ with cylindrical coordinates $(r,\phi,z)$ centered on the star. The gas has density $\rhog$, pressure $\Pg$, and adiabatic index $\gamma$. These yield the squared radial Brunt-V\"{a}is\"{a}l\"{a} frequency, 
\begin{align}
N_r^2 \equiv -\frac{1}{\gamma\rhog }\frac{\p\Pg}{\p r}\frac{\p S}{\p r},\label{Nr2_def}
\end{align}
where $S \equiv \ln{\left(\Pg/\rhog^\gamma\right)}$ is the dimensionless entropy; and the dimensionless radial pressure gradient
\begin{align}
\eta \equiv -\frac{1}{2 r\Omega^2 \rhog}\frac{\p\Pg}{\p r},\label{eta_def}
\end{align}
where $\Omega=\sqrt{GM_*/r^3}$ is the Keplerian frequency and $G$ is the gravitational constant. 
For radially smooth, thin disks in equilibrium, $\left|N_r\right| \sim O(\hgas\Omega)$ and $\left|\eta\right|\sim O(\hgas^2)$, where
\begin{align}
    \hgas \equiv \frac{\Hgas}{r}
\end{align}
is the disk aspect ratio and $\hgas\simeq 0.05$---$0.1$ for PPDs.

The COS requires $N_r^2<0$, together with thermal losses. On the other hand, a dusty gas with $\eta\neq 0$ exhibits dust-gas relative drift, which drives the SI { \citep{youdin05}. For a recent explanation of its instability mechanism, see \cite{magnan24}.} 


\subsection{Local model}

We consider scales much smaller than the typical radius $r$ and thus adopt the shearing box framework \citep{goldreich65} to focus on a small patch of the disk centered at a fiducial radius $r_0$ in the midplane, $(r_0, \phi_0 - \Omega_0 t, 0)$, where $\Omega_0$ is the local Keplerian frequency. For clarity, hereafter, we drop the subscript zero. Cartesian coordinates $(x,y,z)$ in the box correspond to the global disk's radial, azimuthal, and vertical directions.  

We are interested in sub-sonic gas dynamics but wish to retain (radial) buoyancy effects. We thus adopt the Boussinesq shearing box equations derived by \cite{latter17b} and take its unstratified limit. The gas has a solenoidal velocity $\vgtilde$, and we introduce the buoyancy variable $\theta$ (with dimensions of length) that tracks density fluctuations associated with the gas temperature evolution. The gas equations are:

\begin{align}
    \nabla\cdot\vgtilde  = & 0,\\
    \frac{\p\vgtilde}{\p t} + \vgtilde\cdot\nabla\vgtilde = & -\frac{\nabla p}{\rhogref} - 2\Omega\hat{\bm{z}}\times\vgtilde \notag\\ 
    &+ 3\Omega^2 x \hat{\bm{x}}  - N_r^2\theta\xhat + \nu \nabla^2\vgtilde \notag \\
    & + 2\eta r\Omega^2 \hat{\bm{x}} - \frac{\epsilon}{\taus}\left(\vgtilde - \vdtilde\right),\label{gas_mom}\\
    \frac{\p \theta}{\p t} + \vgtilde\cdot\nabla \theta &= \dd\vgtildex - \frac{\theta}{\tcool} + \xi \nabla^2\theta\label{gas_energy},
\end{align}
where $p$ is the local pressure fluctuation. Here, the background density $\rhogref$ is a constant. It will be convenient to refer to the enthalpy { $W$, defined via, 
\begin{align}
    W\equiv \int \frac{dp}{\rhog}.
\end{align}
For $\rhog$ approximately uniform, we have $W=p/\rhog$ plus a constant. Thus $W$ is } equivalent to the pressure in the Boussinesq approximation. We will use `pressure' and `enthalpy' interchangeably. 

In the shearing box, $N_r$ and $\eta$ are constants and correspond to their equilibrium, local values in the global disk. We also take the kinematic gas viscosity $\nu$ to be constant. In Eq. \ref{gas_mom}, the final term $\propto\epsilon$ represents feedback from dust, described below. 

 In the energy equation (\ref{gas_energy}), $\dd \vgtildex$ is the deviation from the equilibrium radial gas velocity, $\tcool$ is a constant, optically thin cooling timescale, and $\xi$ is the constant thermal diffusion coefficient. { We include both thermal loss models here for completeness. See \cite{lyra11} for a similar thermal treatment. However, in practice, we will effectively neglect the optically thin cooling by adopting a long cooling timescale.}   

Following \cite{lehmann23}, we append dust as a pressureless fluid interacting with the gas through a drag force characterized by a constant stopping time $\taus$. The dust has velocity $\vdtilde$ and an associated dust-to-gas mass density ratio $\epsilon$ that evolves according to mass conservation. The dust equations are: 
\begin{align}
    \frac{\p\epsilon}{\p t} + \nabla\cdot\left(\epsilon\vdtilde\right) &= D \nabla ^2 \epsilon \label{dust_mass},\\
    \frac{\p\vdtilde}{\p t} + \vdtilde\cdot\nabla\vdtilde = & - 2\Omega\hat{\bm{z}}\times\vdtilde 
    + 3\Omega^2 x \hat{\bm{x}} \notag\\ & 
    - \frac{1}{\taus}\left(\vdtilde - \vgtilde\right) + \nud\nabla^2\vdtilde.\label{dust_mom}
\end{align}
Here, $D$ and $\nud$ are constant diffusion and viscosity coefficients, respectively. The fluid treatment applies to dust grains tightly coupled to the gas, which have corresponding $\taus\Omega\ll 1$ \citep{jacquet11}. 

In Eqs. \ref{gas_mom} and \ref{dust_mass}---\ref{dust_mom}, we include viscosity and mass diffusion primarily for numerical stability, but one could also motivate these terms by attributing them to some underlying turbulence. For simplicity, we set 
\begin{align}
    D = \nud = \nu
\end{align}
in practice, but will retain separate notations to keep track of their origin. On the other hand, thermal losses or diffusion in Eq. \ref{gas_energy} is a physical requirement for the COS to operate.  
 
\subsection{Gas-based formulation for axisymmetric dynamics} 
Instead of evolving the dust velocity directly, we evolve the relative dust-gas velocity 
\begin{align}
\Dv \equiv \vdtilde - \vgtilde.
\end{align}
The equation for $\Dv$ is obtained by subtracting Eq. \ref{gas_mom} from \ref{dust_mom}.  
It is also convenient to define the gas velocity $\vg$ relative to its equilibrium value in the dust-free limit, such that the total velocity is
\begin{align}
    \vgtilde = \vg - \frac{3}{2}\Omega x \yhat - \eta r \Omega \hat{\bm{y}}
\end{align}

In terms of $\vg$ and $\Dv$ and restricting to axisymmetric flow ($\p_y \equiv 0$), the governing equations become:  
\begin{align}
 &\nabla\cdot\vg  = 0,\label{gas_based_gas}\\
&\frac{\p\vg}{\p t} + \vg\cdot\nabla\vg =  2\Omega\vgy\xhat - \frac{\Omega}{2}\vgx\yhat + \nu\nabla^2\vg \notag \\ 
&\phantom{\frac{\p\vg}{\p t} + \vg\cdot\nabla\vg=} - \nabla W -N_r^2\theta\xhat + \frac{\epsilon}{\taus}\Dv ,\label{gas_based_mom}\\
&\frac{\p \theta}{\p t} + \vg\cdot\nabla \theta = \dd\vgx - \frac{\theta}{\tcool} + \xi\nabla^2\theta,\label{gas_based_energy}\\
&\frac{\p\epsilon}{\p t} + \vg\cdot\nabla\epsilon= - \nabla\cdot\left(\epsilon\Dv\right) + D \nabla^2\epsilon ,\label{gas_based_dg}\\
&\frac{\p \Dv}{\p t} + \left(\vg\cdot\nabla\right)\Dv + \left(\Dv\cdot\nabla\right)\vg + \left(\Dv\cdot\nabla\right)\Dv\notag\\
    & = 2\Omega\Dvy\xhat - \frac{\Omega}{2}\Dvx\yhat + \nud \nabla^2\Dv \notag \\  
    & \phantom{=}+ \nabla W - \frac{(1+\epsilon)}{\taus}\Dv - 2\eta r \Omega^2\xhat + N_r^2\theta\xhat,\label{gas_based_Dv}
\end{align}
where we set $\nu=\nud$ when deriving Eq. \ref{gas_based_Dv}. 

\subsection{Alternative forms of the dust-to-gas ratio equation}

By taking the divergence of Eq. \ref{gas_based_mom} and combining it with Eq. \ref{gas_based_dg}, we obtain an alternative form of the dust-to-gas ratio equation as:
\begin{align}
    \frac{\p\epsilon}{\p t} + \vg\cdot\nabla\epsilon=& \taus\nabla\cdot\left[\left(2\Omega\vgy - N_r^2\theta\right)\xhat - \vg\cdot\nabla\vg\right] \notag \\
    &+D\nabla^2\epsilon -\taus \nabla^2 W .\label{gas_based_dg_alt}
\end{align}
This has the advantage of not involving $\Dv$, and its linearized form is simple. Notice also that the dust-trapping nature of pressure maxima is explicitly reflected in Eq. \ref{gas_based_dg_alt}, which is opposed by dust diffusion.

For numerical simulations, however, it is desirable to ensure that $\epsilon>0$. To this end, we evolve the quantity
\begin{align}
    Q \equiv \epsilon_0 \ln{\left(\frac{\epsilon}{\epsilon_0}\right)},
\end{align}
where $\epsilon_0$ is the initial dust-to-gas ratio and thus $Q=0$ initially. Then $\epsilon = \epsilon_0\exp{\left(Q/\epsilon_0\right)}>0$. 
The equation for $Q$ is 
\begin{align}
    \frac{\p Q}{\p t} =& -\left(\vg + \Dv\right)\cdot\nabla Q - \epsilon_0\nabla\cdot\Dv \notag \\ &+ D\left(\nabla^2 Q + \frac{\left|\nabla Q\right|^2}{\epsilon_0}\right). \label{gas_based_dg_Q}
\end{align}
The term $\propto |\nabla Q|^2$ arises from expressing the diffusion term $\propto \nabla^2\epsilon$ (see Eq. \ref{gas_based_dg}) in terms of $Q$. This formulation was also employed by \cite{wu24}.

\subsection{Equilibrium}

The equilibrium consists of constant velocities
\begin{align}
&\vgx = \frac{2\epsilon\st}{\st^2+(1+\epsilon)^2}\eta r \Omega, \label{eqm_vgx_full}\\
&\vgy = \frac{\epsilon(1+\epsilon)}{\st^2+(1+\epsilon)^2}\eta r \Omega,\\
&\vgz = 0,
\end{align}
and constant differential velocities
\begin{align}
    \Delta v_{x} &= -\frac{2\st(1+\epsilon)}{\st^2+(1+\epsilon)^2}\eta r \Omega,\label{Dvx_eqm}\\
    \Delta v_{y} &= \frac{\st^2}{\st^2+(1+\epsilon)^2}\eta r \Omega,\label{Dvy_eqm}\\
    \Delta v_z   &=0 \label{eqm_Dvz_full}. 
\end{align}
The equilibrium pressure and buoyancy variables are zero,
\begin{align}
 W = \theta = 0. 
\end{align}

Strictly speaking, a dusty, non-isothermal gas disk cannot remain in thermodynamic equilibrium: the dust-induced radial gas flow would transport the background entropy and act as an effective heat source. Our local model neglects this by assuming a constant heat sink exists to offset it, such that $\dd \vgx $ is the perturbation relative to the background gas drift. See \S 6.6.1 of \citet{lehmann23} for a further discussion on this issue.

\subsection{Dimensionless parameters { and fiducial values}}\label{parameters}
We define several dimensionless parameters to label our simulations. Although the Boussinesq shearing box is unaware of  $\Hgas$ or $\hgas$, we use these in the definitions below to connect our parameters to the global disk.    

The buoyancy parameter $\nrsq$ describes the radial stratification,
\begin{align}
    \nrsq \equiv -\frac{N_r^2}{\Omega^2},
\end{align}
so that radially buoyant disks have $\nrsq>0$. { The fiducial $\nrsq=0.1$.} 

We define the reduced radial pressure gradient parameter 
\begin{align}
    \Pi \equiv \frac{\eta}{\hgas} 
\end{align}
to set $\eta$ in the box. In the linear theory of the classical SI, $\Pi$ is the relevant parameter instead of $\eta$ itself. { Our fiducial setup adopts $\Pi=0$.}

Note that, in a global disk, $\Pi$ and $\nrsq$ are not independent since { they are related by the radial entropy gradient.} Thus, one cannot vanish without the other. However, the local model treats them as separate parameters representing different physical effects. 

Thermal diffusion is parameterized by the P\'{e}clet number,
\begin{align}
    \peclet \equiv \frac{\Hgas^2\Omega}{\xi}.
\end{align}
{ The fiducial $\peclet = 160\pi^2$.} We effectively disable Newtonian cooling by setting $\beta\equiv \tcool\Omega = 10^6$. 

We characterize dissipative effects by the Reynolds number 
\begin{align}
    \reynolds = \frac{\Hgas^2\Omega}{\nu},
\end{align}
and likewise for $\nud$ and $D$. With this definition, $\reynolds^{-1}$ is equivalent to the {  $\alpha$ parameter} often used to parameterize turbulent angular momentum transport in classical viscous accretion disks \citep{shakura73,pringle81}. { We fix $\reynolds=10^{5}$ ($\alpha = 10^{-5}$) for all calculations and runs.}

The Stokes number $\st$ characterizes the degree of dust-gas coupling,
\begin{align}
    \st \equiv \taus\Omega. 
\end{align} 
{ Our fiducial $\st=0.1$. This is relatively large compared to the smallest pebbles expected in PPDs \citep[which have $\st=10^{-3}$ to $\st=10^{-2}$, ][]{ormel24}. The fluid treatment of dust requires $\taus\omega_\mathrm{f}\ll 1$, where $\omega_\mathrm{f}$ is the characteristic frequency of the gas dynamics. The fluid approximation is expected to remain applicable since the COS corresponds to inertial waves, which have $\omega_\mathrm{f}\leq \Omega$. 
}

Where necessary, we denote the initial dust-to-gas ratio by $\epsilon_0$, while we omit the subscript zero in discussing equilibrium solutions and linear theory for clarity. { The fiducial $\epsilon_0=0.01$.} 

\subsection{Disabling dust feedback}\label{no_feedback}
We can disable dust feedback on the gas as follows: 
We set the drag term in the gas momentum equation (\ref{gas_based_mom}) to zero, $\epsilon\Dv/\taus\to 0$.  Similarly, the drag term in the differential velocity equation (\ref{gas_based_Dv}) becomes $(1+\epsilon)\Dv/\taus \to \Dv/\taus$. 

Disabling feedback allows one to examine how the gas flow and drag forces affect dust concentration and dispersal without complications from drag instabilities such as the SI. Neglecting feedback is usually justified for $\epsilon\ll 1$, but we shall find that it affects the COS even in this regime.

\subsection{Terminal velocity approximation}
One expects $\Dv$ to be small for tightly coupled grains and formally vanish for perfectly coupled dust. This leads to an approximate but explicit expression for $\Dv$ in the `terminal velocity approximation' (TVA). The governing equations are then reduced by one at the expense of increased complexity of the drag term in the gas momentum equation and the dust continuity equation, which also involves $\Dv$. 
The TVA is described in Appendix \ref{TVA} and is used for 
interpreting some results obtained from the full treatment. 
\section{Linear theory}\label{linear}
We first show that the above system of equations encapsulates the COS and the SI. We consider axisymmetric, Eulerian perturbations of the form 
\begin{align}
    \dd \vg =\operatorname{Re}\left[\widehat{\dd \vg} \exp{\left(\sigma t + \ii k_x x + \ii k_z z\right)}\right],\label{linear_pert_form}
\end{align}
and similarly for other variables; where $\widehat{\phantom{s}}$ denotes a complex amplitude, $\sigma$ is the complex growth rate, and $k_{x,z}$ are real wavenumbers. We take $k_{x,z}>0$ without loss of generality. For clarity, hereafter we drop the $\widehat{\phantom{s}}$. 

The linearized equations are: 
\begin{align}
&\ik\cdot\dd\vg = 0,\label{linear_div}\\
&\sigma \dd\vg + \ikx \vgx \dd\vg = 2\Omega\dd\vgy\xhat 
- \frac{\Omega}{2}\dd\vgx\yhat  - \nu k^2\dd\vg  \notag \\ 
&\phantom{\sigma \dd\vg + \ikx \vgx \dd\vg =} 
-\ik \dd W - N_r^2\dd\theta\xhat \notag\\
&\phantom{\sigma \dd\vg + \ikx \vgx \dd\vg =} 
+\frac{1}{\taus}\left(\dd\epsilon\Dv + \epsilon\dd\Dv\right),\label{linear_vg}\\
&\sigma\dd\theta + \ikx\vgx\dd\theta = \dd\vgx - \frac{\dd\theta}{\tcool} - \xi k^2 \dd\theta,\\
&\sigma\dd\epsilon + \ikx\vgx\dd\epsilon = \ikx \taus\left(2\Omega\dd\vgy - N_r^2\dd\theta\right) \notag \\
&\phantom{\sigma\dd\epsilon + \ikx\vgx\dd\epsilon =}
+ \taus k^2 \dd W - Dk^2\dd\epsilon,\label{linear_dg}\\
 &\sigma\dd\Dv +  \ikx\vgx\dd\Dv + \ikx\Dvx\dd\vg + \ikx\Dvx \dd\Dv \notag\\
    &= 2\Omega\dd\Dvy\xhat - \frac{\Omega}{2}\dd\Dvx\yhat - \nud k^2 \dd\Dv. \notag \\  
    &\phantom{=} + \ik \dd W -\frac{\Dv}{\taus}\dd\epsilon - \frac{(1+\epsilon)}{\taus}\dd\Dv + N_r^2\dd\theta\xhat.\label{linear_Dv} 
\end{align}
Note that we used the alternative form of the dust-to-gas ratio equation (Eq. \ref{gas_based_dg_alt}) to obtain Eq. \ref{linear_dg}. The above system constitutes a generalized eigenvalue problem of the form $\bm{A}\bm{U} = \sigma \bm{B}\bm{U}$; where the elements of the matrices $\bm{A}$ and $\bm{B}$ can be read off Eqs. \ref{linear_div}---\ref{linear_Dv}, and $\bm{U} \equiv \left[\dd W, \dd\epsilon, \dd\vg, \dd\Dv, \dd\theta \right]^T$ is the 9-element eigenvector. It is possible to reduce the linear problem to a standard eigenvalue calculation by using the incompressibility condition to eliminate the pressure perturbation.  

In the discussion below, we refer to the dimensionless wavenumber 
\begin{align}
    K_{x,z} \equiv k_{x,z}\Hgas.
\end{align}

\subsection{Convective overstability}\label{COS_linear}

Our fiducial parameter values for the COS are $\nrsq=0.1$, $\peclet=160\pi^2$, and $\reynolds=10^5$. We neglect the background radial pressure gradient by setting $\Pi=0$, suppressing the SI. 

\begin{figure*}
    \centering
    \includegraphics[scale=0.67,clip=true, trim=0cm 0cm 3cm 0cm]{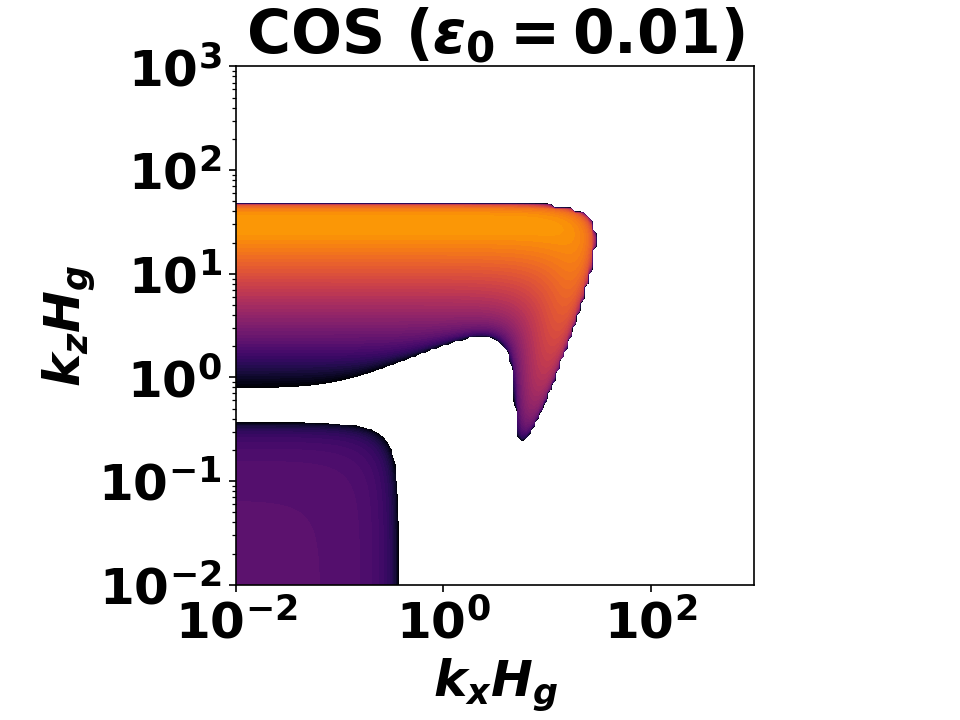}\includegraphics[scale=0.67,clip=true, trim=2.5cm 0cm 0cm 0cm]{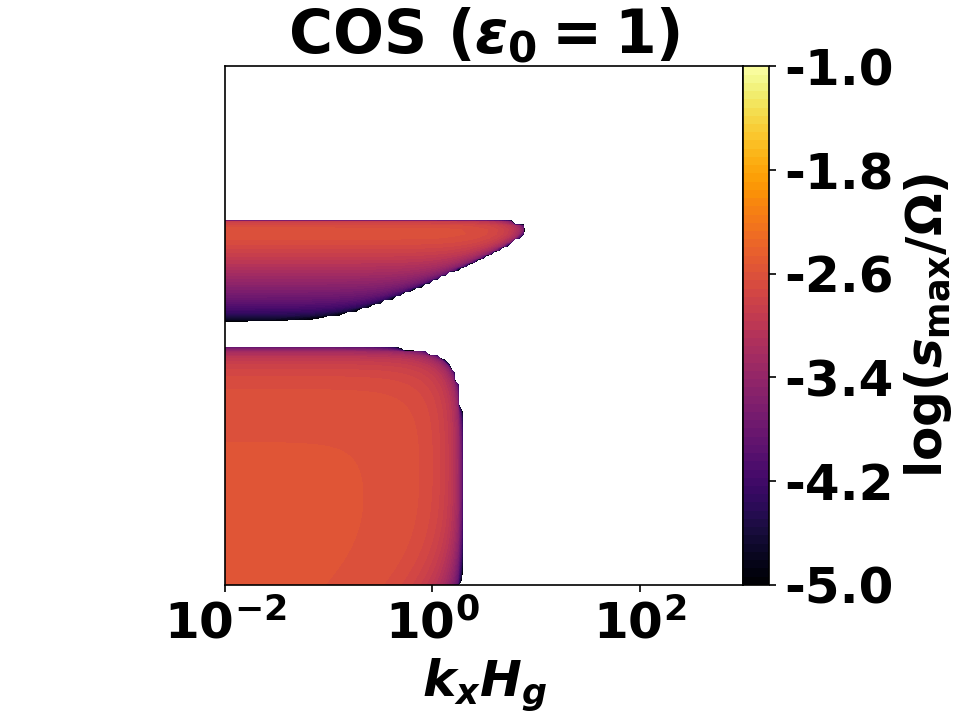}
    \caption{COS with $\epsilon_0=0.01$ (left) and $\epsilon_0=1$ (right). Other parameters are: $\nrsq=0.1$, $\peclet=160\pi^2$, $\reynolds=10^5$, and $\st=0.1$. No background global radial pressure gradient is applied.}
    \label{COS_exact}
\end{figure*}

Fig. \ref{COS_exact} show the maximum growth rates,  
$s_\mathrm{max}$, as a function of $K_{x,z}$ for $\epsinit =0.01$ (left) and $\epsinit=1$ (right). In either case, COS corresponds to the higher-$K_z$, nearly horizontal `slab' of modes. The `square' at small $K_{x,z}$ are destabilized cooling modes in a dusty gas, as discussed by \cite{lehmann23}. These require long cooling times, which apply to long wavelengths under thermal diffusion. However, they are irrelevant to this work because they have smaller growth rates, scales larger than our simulation domains, or both.  

For $\epsilon_0=0.01$, the largest COS growth rate is $s_\mathrm{max}\simeq 0.013\Omega$, occurring at $K_z \sim 40 \simeq \sqrt{\peclet}$, in agreement with \citetalias{teed21}. For $\epsilon_0=1$, the COS is weakened to $s_\mathrm{max}\simeq 0.003\Omega$. Dust loading reduces COS growth rates because the effective squared buoyancy frequency is reduced by $\nrsq \to \nrsq/(1+\epsilon_0)$. 
Modes are suppressed by viscosity at high wavenumbers. According to \cite{latter16}, the cut-off $K_z\propto (\nrsq/\operatorname{Pr})^\frac{1}{4}$ for small Prandtl numbers $\operatorname{Pr}\equiv \nu/\xi\ll 1$. Consequently, there is a slight reduction in the cut-off vertical wavenumber as $\epsilon_0$ increases as it decreases the effective $\nrsq$, as observed. At large $\epsilon_0$, finite drag forces further reduce the cut-off radial wavenumber, an effect explained in \cite{lehmann23}.




\subsection{No dust concentration by channel modes}\label{no_dust_concen_channel}
We can assess the ability of linear gas modes to concentrate dust as follows. For this discussion, we neglect viscosity, diffusion, and a background radial dust drift. We also neglect dust feedback as described in \S\ref{no_feedback}. We linearize the primitive dust-to-gas ratio equation (\ref{gas_based_dg}) and apply the TVA to find 
\begin{align*}
\sigma \dd\epsilon &= -\epsilon\taus\left(\ikx N_r^2\dd\theta - k^2 \dd W\right),\\
&=-2\ikx \epsilon\st\dd\vgy,
\end{align*}
where we used the divergence of the gas momentum equation (neglecting dust drag) for the second equality. 

Channel modes with $k_x=0$ cannot concentrate dust. Since the most unstable gaseous COS modes have $k_x\to 0$ \citep{lyra14,lehmann23}, the COS cannot drive meaningful dust concentrations in the linear regime. This is indeed what we observe in simulations. Thus, one must examine the nonlinear regime. 

\subsection{Streaming instability}
\begin{figure}
    \centering
    \includegraphics[width=\linewidth]{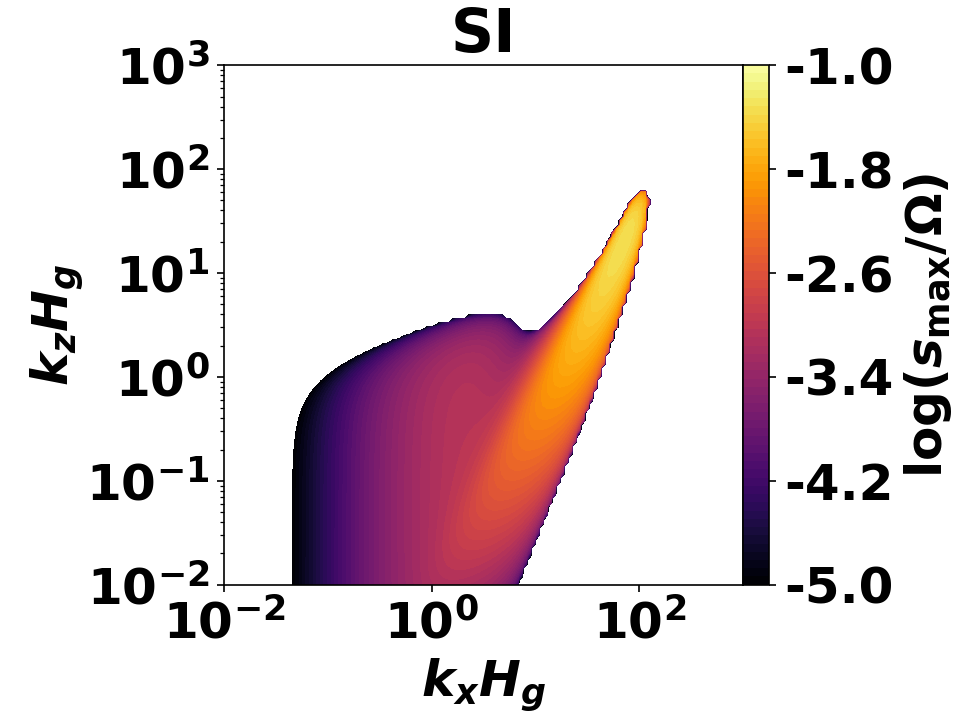}
    \caption{SI with $\epsilon_0=3$ and a background pressure gradient with $\Pi=0.1$. Other parameters are: $\peclet=160\pi^2$ $\nrsq=-0.1$, $\reynolds=10^5$, $\st=0.1$.}
    \label{SI_exact}
\end{figure}

We also recover the SI, powered by a background dust-gas relative drift, by setting $\Pi=0.1$. We suppress the COS by taking $\nrsq=-0.1$. Other parameters are as above. We consider $\epsilon_0=3$ to obtain appreciable growth rates. Fig. \ref{SI_exact} shows growth rates as a function of $K_{x,z}$. Apart from high $K_{x,z}$ modes being stabilized by viscosity, this plot is qualitatively similar to the classical SI in isothermal disks \citep[e.g.][see their Fig. 2]{lin22}. We checked that the most unstable SI modes are unaffected by $\peclet$. 


\section{Numerical simulations}\label{numerical}




We use the \textsc{dedalus} spectral code\footnote{\url{https://dedalus-project.org/}} \citep{burns19} to evolve the gas equations (\ref{gas_based_gas}---\ref{gas_based_energy}), the full relative drift equation (\ref{gas_based_Dv}), and the positive-definite formulation of the dust-to-gas ratio equation (\ref{gas_based_dg_Q}). The solenoidal condition (Eq. \ref{gas_based_gas}) is supplemented with a pressure gauge following the \textsc{dedalus} documentation\footnote{\url{https://dedalus-project.readthedocs.io/en/latest/pages/gauge_conditions.html}}. Axisymmetry ($\p_y\equiv0$) is assumed throughout.

We set the radial and vertical domain sizes to $L_x = \Hgas$ and $L_z=\Hgas/2$, respectively, and adopt a resolution of $N_x\times N_z = 2048\times1024$, which is justified in Appendix \ref{resolution}. For axisymmetric flow, the boundary conditions are strictly periodic.  We use a standard dealiasing factor of $3/2$ and the RK443 time stepper in \textsc{dedalus}. We use a Courant–Friedrichs–Lewy (CFL) number of $0.4$. We limit the maximum step size to $\taus$ { since} the non-linear drag terms are treated explicitly. 


We adopt units such that $\Hgas=\Omega=1$. Then $P \equiv 2\pi$ corresponds to one orbit. The background gas density $\rhogref=1$. 


\section{Results}\label{results}

\subsection{A fiducial case}\label{result_fid}
Our fiducial simulation adopts the same parameters as the disk described in \S\ref{COS_linear} and is initialized with $\epsilon_0=0.01$ (left panel of Fig. \ref{COS_exact}). This case produces steady zonal flows, making studying dust-trapping by COS-induced pressure bumps easier. Cases with faster thermal diffusion using $\peclet=16\pi^2$, which exhibit more time variability, are presented in \S\ref{pe16pi2}. 

\begin{figure}
    \centering
    \includegraphics[width=\linewidth]{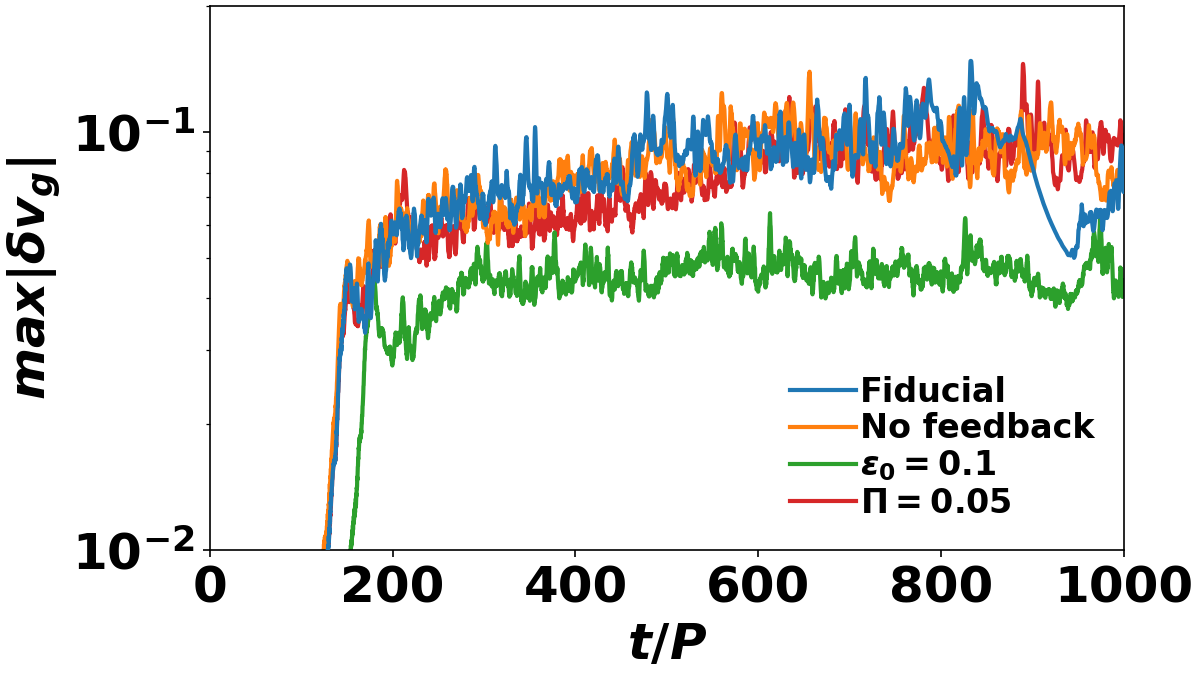}
    \caption{Evolution of the gas velocity perturbations for the COS in the fiducial run (blue), that without feedback (orange), a larger dust-to-gas ratio (green), and with a non-zero pressure gradient (red). 
    }
    \label{COS_vg_Pe1600}
\end{figure}

The blue curve in Fig. \ref{COS_vg_Pe1600} shows the evolution of the maximum gas velocity perturbations for the fiducial run. The orange, green, and red curves correspond to the same run without feedback, with a larger dust-to-gas ratio of $\epsilon_0=0.1$, and a non-zero pressure gradient $\Pi=0.05$, respectively. These additional runs are discussed in the following sections. To aid visualization, we perform a running-time average over $10$ orbits. 

The fiducial run proceeds with a linear phase with a growth rate of $1.15\times10^{-2}\Omega$, slightly smaller than the maximum value expected from linear theory ($1.28\times10^{-2}\Omega$). This may be due to the finite radial domain size, which cannot accommodate the most unstable COS modes with vanishing $k_x$. The run saturates with $\mathrm{max}\left|\delta v_g\right|\sim 0.1$. 

\begin{figure}
    \centering
    \includegraphics[width=\linewidth]{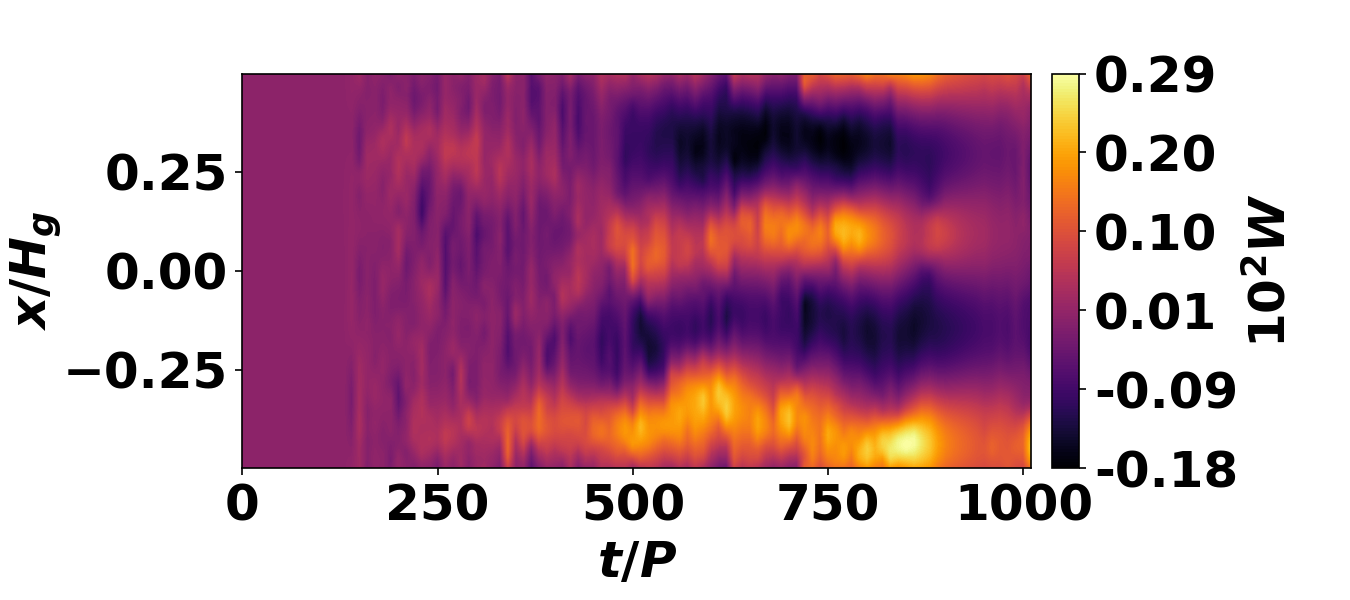}
    \caption{Space-time plot of the vertically-averaged pressure  distribution in the fiducial run.
    }
    \label{COS_spacetime_W_Pe1600}
\end{figure}

\begin{figure}
    \centering
    \includegraphics[width=\linewidth]{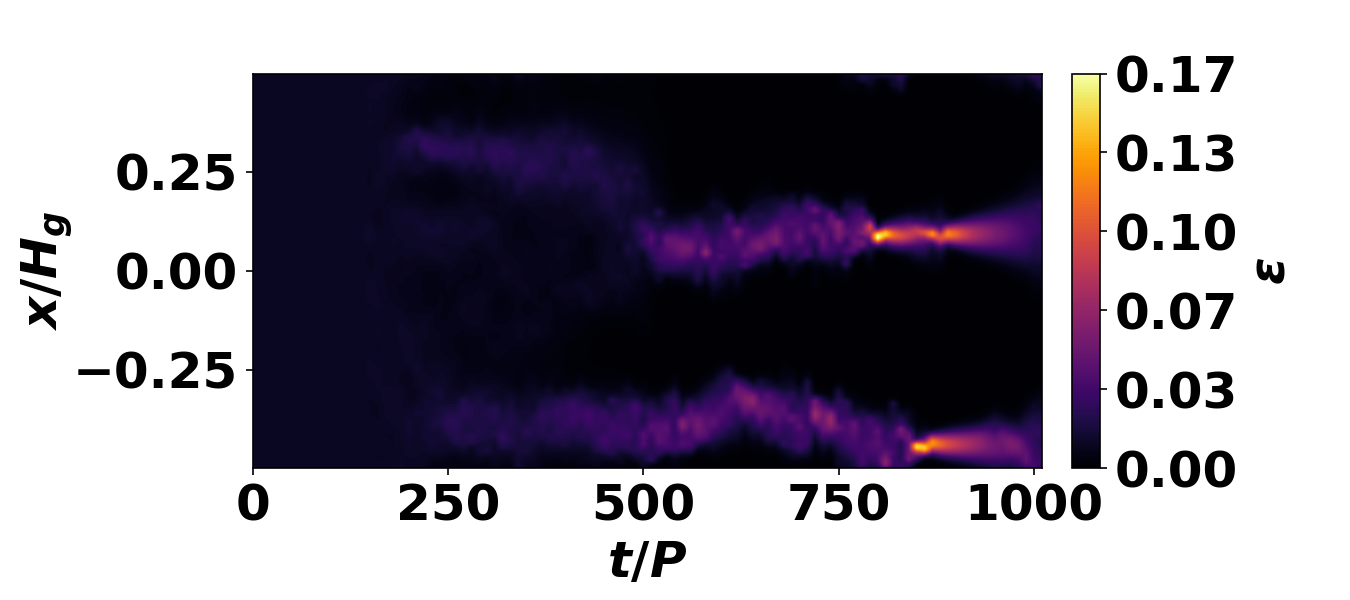}
    \caption{Similar to Fig. \ref{COS_spacetime_W_Pe1600} but for the dust-to-gas ratio. 
    }
    \label{COS_spacetime_eps_Pe1600}
\end{figure}

Figs. \ref{COS_spacetime_W_Pe1600}---\ref{COS_spacetime_eps_Pe1600} show the evolution of the vertically-averaged pressure and dust-to-gas ratio for the fiducial run. The system develops and sustains two quasi-steady pressure rings at $x\simeq -0.4\Hgas$ and $0.1\Hgas$ from $\sim 500$---$800P$, which traps dust at their respective radii. However, the pressure bumps and dust concentrations weaken once $\epsilon$ reaches $O(0.1)$ around $800P$ due to dust feedback onto the zonal flows. 

\begin{figure}
    \centering
    \includegraphics[scale=0.4,clip=true,trim=0cm 1.8cm 0cm 0.8cm]{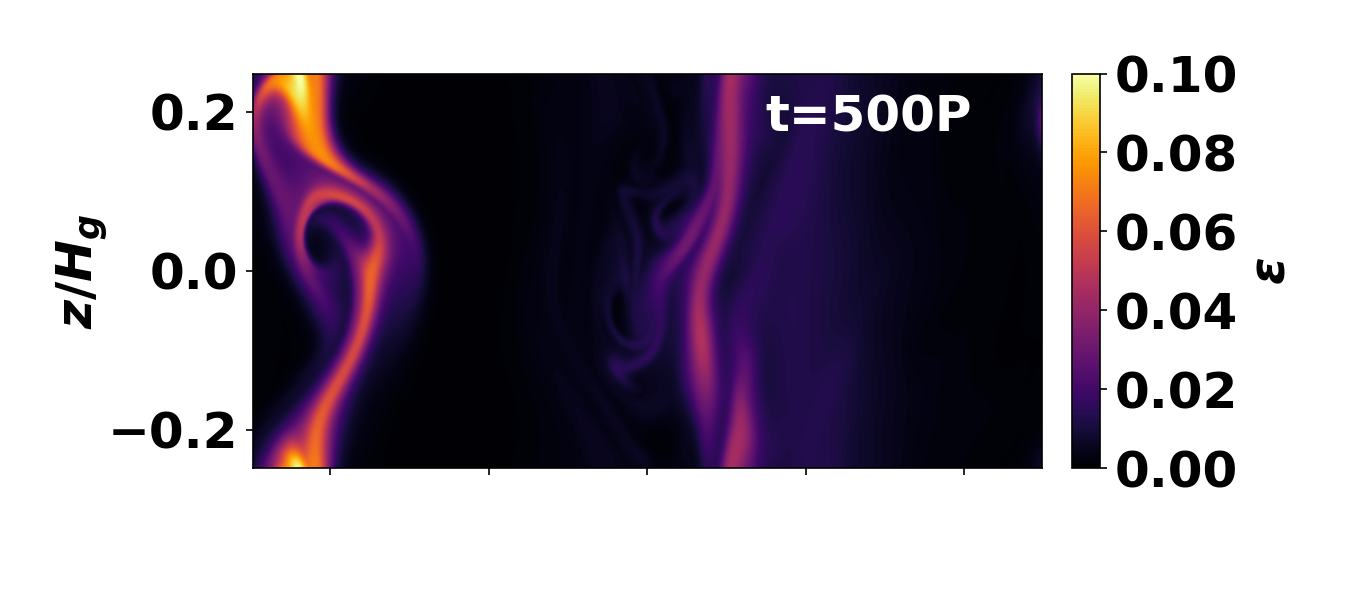}\\
    \includegraphics[scale=0.4,clip=true,trim=0cm 1.8cm 0cm 0.8cm]{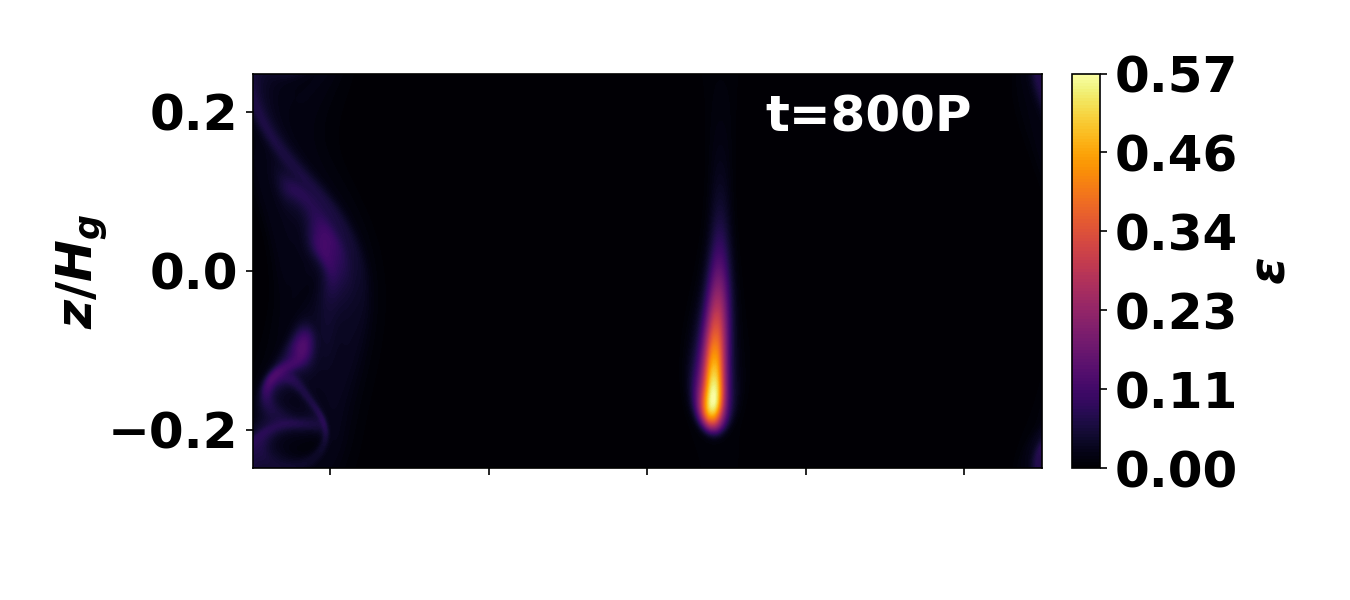}\\
    \includegraphics[scale=0.4,clip=true,trim=0cm 1.8cm 0cm 0.8cm]{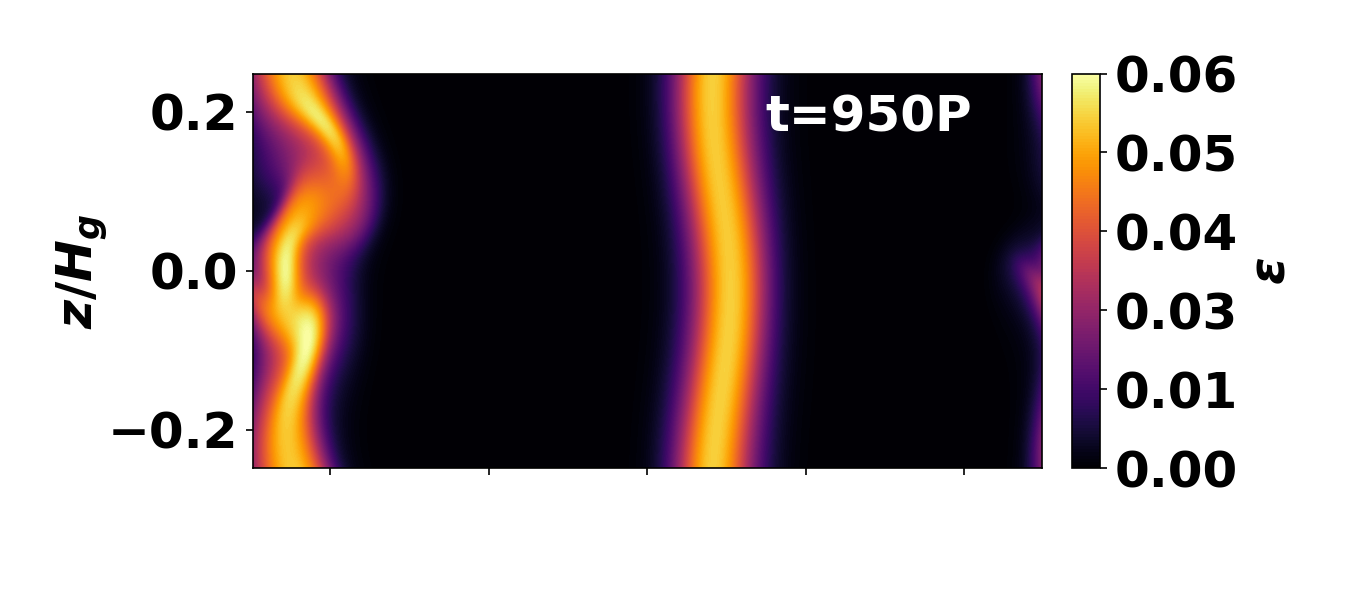}\\
    \includegraphics[scale=0.4,clip=true,trim=0cm 0cm 0cm 0.8cm]{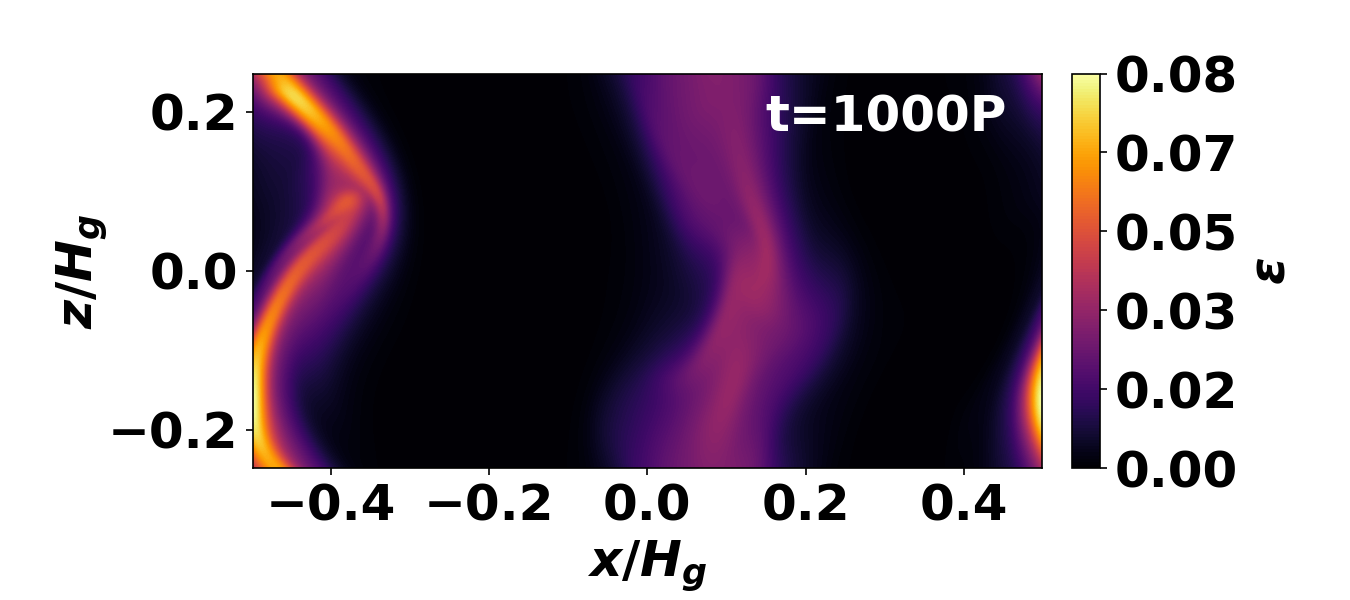}\\
    \caption{Selected snapshots of the dust-to-gas ratios of the fiducial run.}
    \label{epsilon_Pe1600}
\end{figure}

The dust distribution displays significant spatial and time variability within the pressure bumps. This is shown in Fig. \ref{epsilon_Pe1600} with several snapshots of $\epsilon$. Dust-to-gas ratios are typically $\sim 0.1$ within the rings, though it can reach $\sim 0.6$ temporarily. Dust rings are rarely columnar as they frequently undergo regular buckling due to the meridional flows of the COS. { Although the system lacks vertical gravity, vertical flows arise from parasitic inertial waves with nonzero radial wavenumbers and pressure perturbations (\citealt{latter16}, \citetalias{teed21}).}

\begin{figure}
    \centering
    \includegraphics[width=\linewidth]{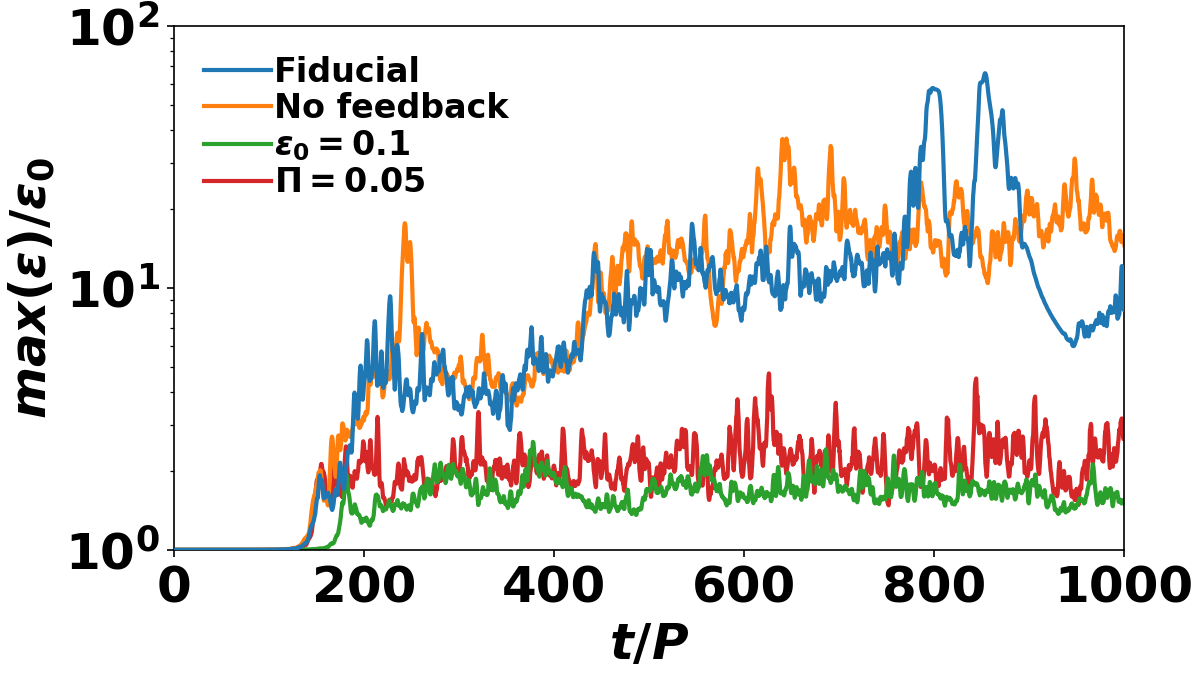}
    \caption{Similar to Fig. \ref{COS_vg_Pe1600} but for the maximum dust-to-gas ratios, normalized by its initial value.
    }
    \label{COS_eps_Pe1600}
\end{figure}

In Fig. \ref{COS_eps_Pe1600}, we plot the maximum $\epsilon$ evolution. We normalize the curves by the initial $\epsilon$ to quantify the ability of COS-induced zonal flows to concentrate dust. In the fiducial case, concentration factors are typically $O(10)$ and appear limited by $\epsilon=0.6$ (attained at $800P$ and $850P$) as $\epsilon$ rapidly declines afterward. 

\subsection{Effect of dust feedback}

 We examine the role of dust feedback with one run strictly without feedback and one run with stronger feedback using $\epsilon_0=0.1$. These are shown as the orange and green curves in Figs. \ref{COS_vg_Pe1600}---\ref{COS_eps_Pe1600}, respectively. 

COS turbulence is weakened by dust feedback for $\epsilon\gtrsim O(0.1)$. This is evident from Fig. \ref{COS_vg_Pe1600}, which shows that with $\epsilon_0=0.1$, $\operatorname{max}\left|\delta\vg\right|$ decreases by a factor of $\sim2$ compared to the fiducial case. On the other hand, the run without feedback behaves similarly to the fiducial case until $900P$, whence $\epsilon$ reaches $O(0.1)$, and activity drops towards the $\epsilon_0=0.1$ run. 

Fig. \ref{COS_eps_Pe1600} shows that even without feedback, dust concentrations are limited to a factor of $O(10)$, due to the internal turbulence of zonal flows. The two epochs of rapid dust growth at $800P$ and $850P$ in the fiducial run show that feedback may temporarily boost concentrations, probably via streaming-type instabilities. However, for the most part, feedback weakens dust concentrations: the fiducial run has marginally lower concentrations than the run without feedback. 

\begin{figure}
    \centering
    \includegraphics[scale=0.2,clip=true,trim=4cm 0cm 2cm 0.0cm]{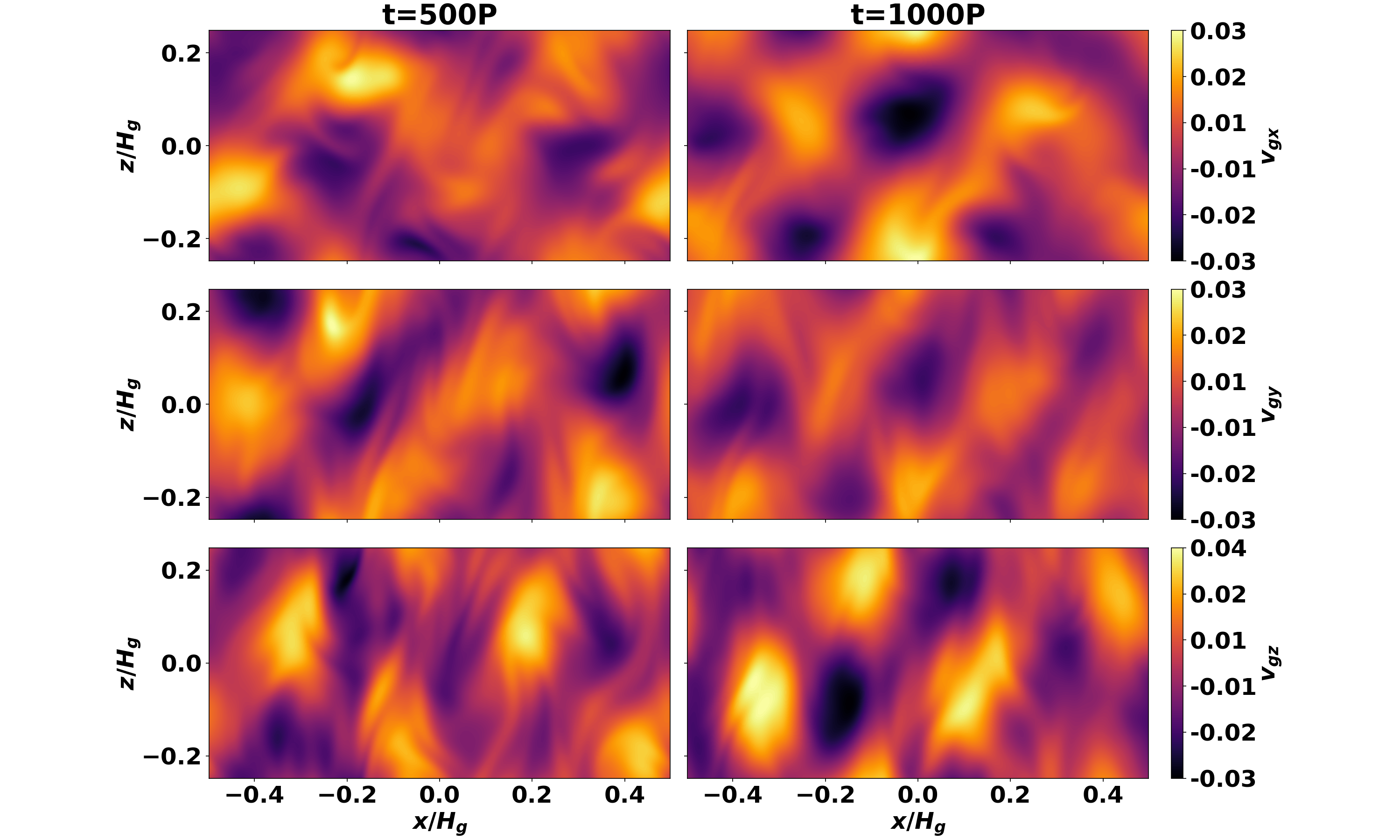}
    \caption{Wave turbulence exhibited by the run with stronger feedback using $\epsilon_0=0.1$. The gas velocity components are shown.}
    \label{epsilon_Pe1600_eps0.1_vg}
\end{figure}

\begin{figure}
    \centering
    \includegraphics[width=\linewidth]{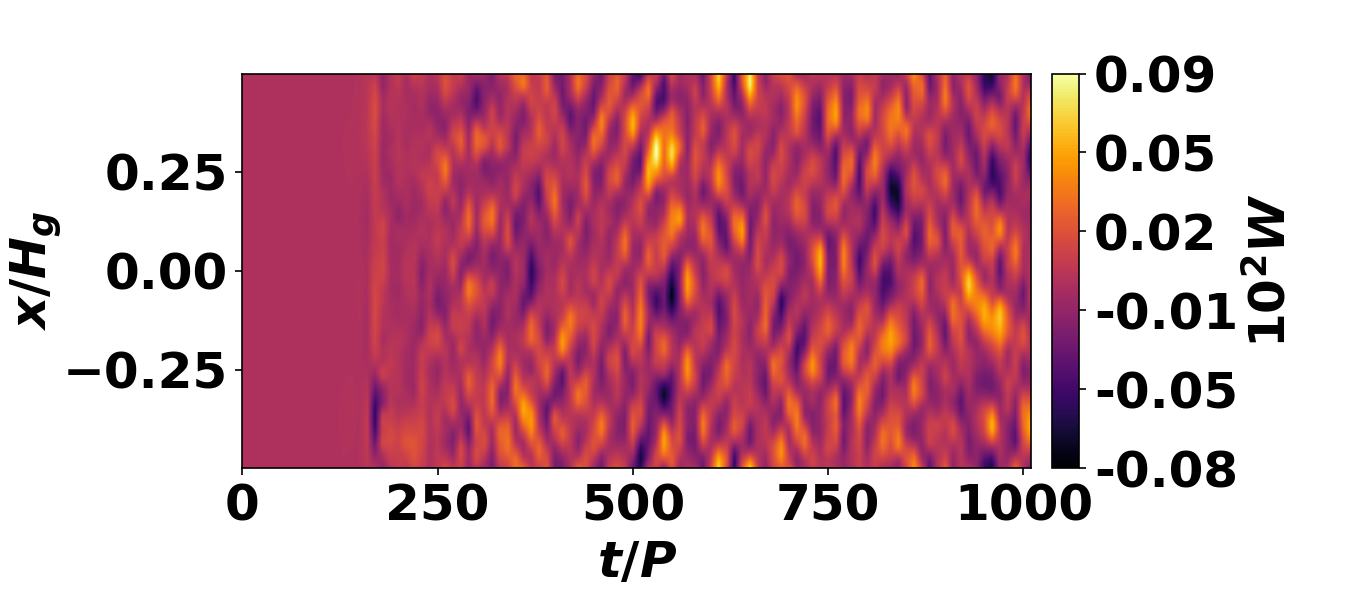}
    \caption{Space-time plot of the vertically-averaged pressure distribution for the run with stronger dust feedback using $\epsilon_0=0.1$ (green curves in Figs. \ref{COS_vg_Pe1600} and \ref{COS_eps_Pe1600}). No zonal flows form in this case.  
    }
    \label{COS_spacetime_W_Pe1600_eps0.1}
\end{figure}

For $\epsilon_0=0.1$, dust concentrations are significantly reduced to less than a factor of two. Here, we find a qualitatively different reason: zonal flows do not form. Instead, the system remains in `wave turbulence' \citepalias{teed21}. This is shown in Fig. \ref{epsilon_Pe1600_eps0.1_vg} as snapshots of the velocity fields, which can be compared to Fig. 7 in \citetalias{teed21}. The lack of persistent $\vgy$ perturbations, i.e., zonal flows, produces negligible dust concentrations. This is evident in the space-time pressure evolution in Fig. \ref{COS_spacetime_W_Pe1600_eps0.1}.

\subsection{Angular momentum fluxes}

To interpret the above result, we examine the total turbulent angular momentum flux (AMF), defined as $F\equiv \Fg + \epsilon_0 \Fd$, where 
\begin{align}
    F_\mathrm{g} \equiv \dd \vgx \dd \vgy, \,  F_\mathrm{d} \equiv \dd \vdx \dd \vdy
\end{align}
are the specific AMFs associated with gas and dust, respectively. For the dust, we calculate $\dd \vdx = \dd \vgx + \dd \Delta v_x$, and similarly for $\dd \vdy$. Recall the $\delta$'s denote deviations from the equilibrium values, which are zero in the present case due to the absence of a global pressure gradient.  

\begin{figure}
    \centering
    \includegraphics[scale=0.4,clip=true,trim=0cm 0cm 0cm 0.0cm]{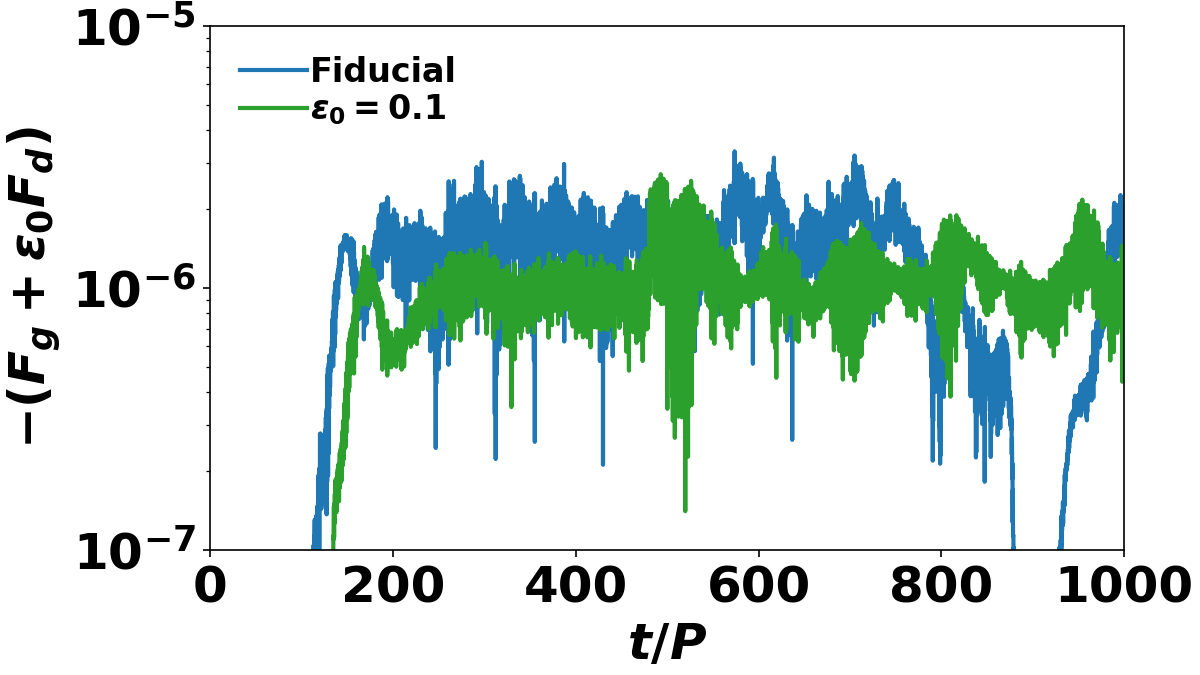}
    \caption{Total angular momentum flux in the fiducial run (blue) and the run with stronger feedback using $\epsilon_0=0.1$. Fluxes are averaged over 100 orbit intervals to improve visibility.}
    \label{amfluxtot_Pe1600_eps0.1}
\end{figure}

In Fig. \ref{amfluxtot_Pe1600_eps0.1}, we compare the box-averaged total AMF between the fiducial case and the $\epsilon_0=0.1$ case. We plot $-F$ since $F<0$ for COS-driven turbulence in the dust-free limit \citepalias{teed21}, corresponding to inwards angular momentum transport. Indeed, $F$ is negative on average, although it has large fluctuations on orbital timescales, sometimes rendering $F>0$ momentarily. 

We find that, on average, $F$ becomes less negative with higher dust abundance. Notice also the significant drop in $|F|$ in the fiducial run at $900P$, coincident with the dispersal of pressure bumps after sufficient dust accumulation. These observations are consistent with \citetalias{teed21}'s explanation for zonal flow formation that involves $F<0$, see also \S\ref{discuss_AMF}.

Moreover, we find $\Fg<0$ and are comparable between the two cases before $800P$, while $\Fd>0$ is marginally larger in the fiducial run. This is offset by the ten-fold increase in dust abundance for $\epsilon_0=0.1$, leading to a noticeably reduced inwards AMF for the mixture. Although $|F|$ only decreases slightly, this appears sufficient to suppress zonal flows. 



\subsection{Effect of a background pressure gradient}

We next introduce a radial pressure gradient by setting $\Pi =  0.05$, which produces a background dust-gas radial drift. The SI is then formally active, but we verified it has lower growth rates than COS modes. (SI modes also have wavelengths exceeding the domain size as the imposed dissipation suppresses smaller-scale modes.) Note that having both $\Pi$ and $\nrsq$ being nonzero is a more self-consistent treatment in the context of global disks; see Eqs. \ref{Nr2_def}---\ref{eta_def}. 

The red curves in Figs. \ref{COS_vg_Pe1600} and \ref{COS_eps_Pe1600} show that, while a radial pressure gradient does not affect the COS-turbulence levels, dust concentrations are significantly reduced to a factor of two, similar to $\epsilon_0=0.1$. Unlike that case, which does not form zonal flows (Fig. \ref{COS_spacetime_W_Pe1600_eps0.1}), here we find zonal flows still form, but dust does not accumulate effectively.



\begin{figure}
    \centering
    \includegraphics[width=\linewidth]{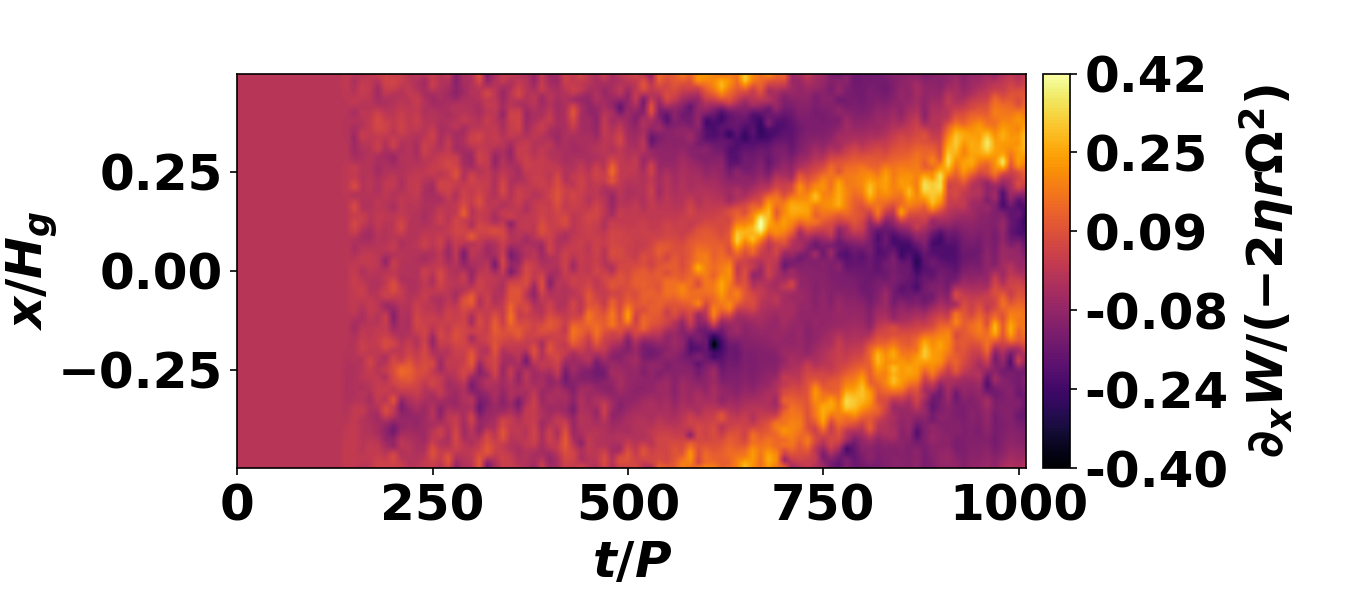}
    \caption{Space-time plot of the vertically-averaged, local radial pressure gradient for the run with a background global radial pressure gradient $\Pi=0.05$ (red curves in Figs. \ref{COS_vg_Pe1600} and \ref{COS_eps_Pe1600}).  
    }
    \label{COS_spacetime_dWdx_Pe1600_eta0.05}
\end{figure}

In Fig. \ref{COS_spacetime_dWdx_Pe1600_eta0.05}, we plot the space-time evolution of the radial pressure gradient ($\p_x W$), which attempts to concentrate dust into zonal flows. We normalize it with a global pressure gradient ($-2\eta r \Omega^2$) that tends to drive a box-wide inward drift. The two contributions are comparable in magnitude around pressure bumps. Beyond the pressure maximum, they work in tandem, but interior to it, the background drift opposes dust accumulation. 

When $\Pi\neq0$, dust concentration becomes more difficult because it drifts in response to the box-wide pressure gradient. Zonal flows are no longer perfect pressure maxima wherein radial drift halts. In the context of a global disk, these correspond to `traffic jams' where dust only slows down but eventually drift through. \citep{pinilla17}. 




\subsection{Intermittent zonal flows with faster thermal diffusion}\label{pe16pi2}

We briefly present selected simulations with $\peclet = 16\pi^2$. In this case, the most unstable vertical wavelength, equal to $2\pi\Hgas \peclet^{-\frac{1}{2}}$ \citepalias{teed21}, is $\Hgas/2$, i.e., the vertical box size, which optimizes the instability. Otherwise, the setup is identical to the fiducial one with $\Pi=0$. 

\begin{figure}
    \centering
    \includegraphics[width=\linewidth]{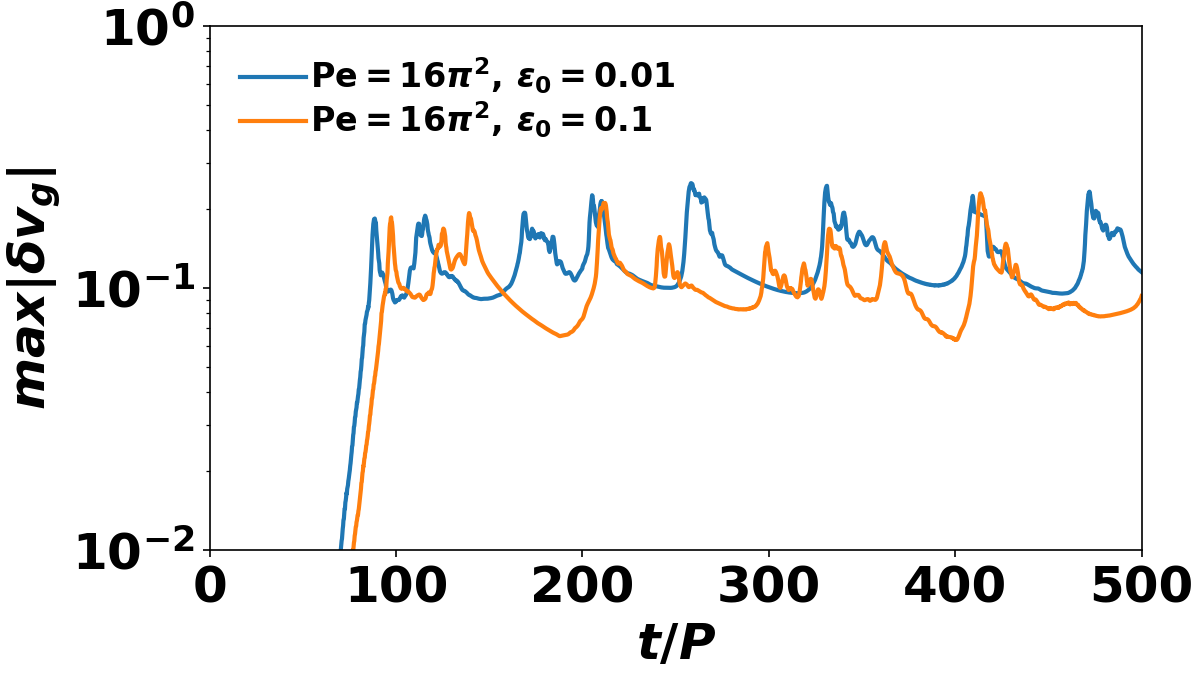}
    \caption{Evolution of the gas velocity perturbations for the COS under stronger thermal diffusion $\peclet=16\pi^2$ for $\epsilon_0=0.01$ (blue) and $\epsilon_0=0.1$ (orange). 
    }
    \label{COS_vg_Pe16pi2}
\end{figure}

Fig. \ref{COS_vg_Pe16pi2} shows the maximum gas velocity perturbations for $\epsilon_0=0.01$ and $0.1$. Here, zonal flows form in both cases, indicating that a stronger instability can offset the stabilization of zonal flow formation by dust feedback. The average amplitudes are slightly larger than the runs with higher $\peclet$. However, they exhibit predator-prey cycles associated with the formation and destruction of zonal flows by the primary COS and secondary parasitic instabilities, respectively. 

\begin{figure}
    \centering
    \includegraphics[width=\linewidth,clip=true,trim=0cm 1.9cm 0cm 0cm]{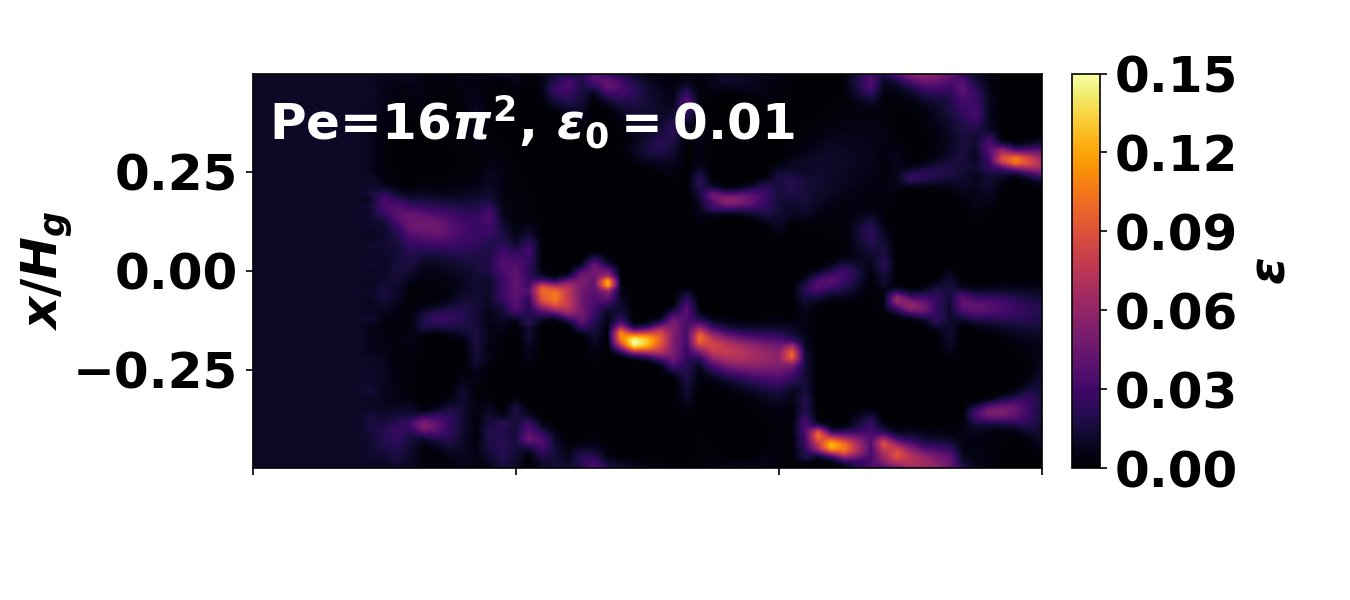}\\
    \includegraphics[width=\linewidth,clip=true,trim=0cm 0cm 0cm 0.7cm]{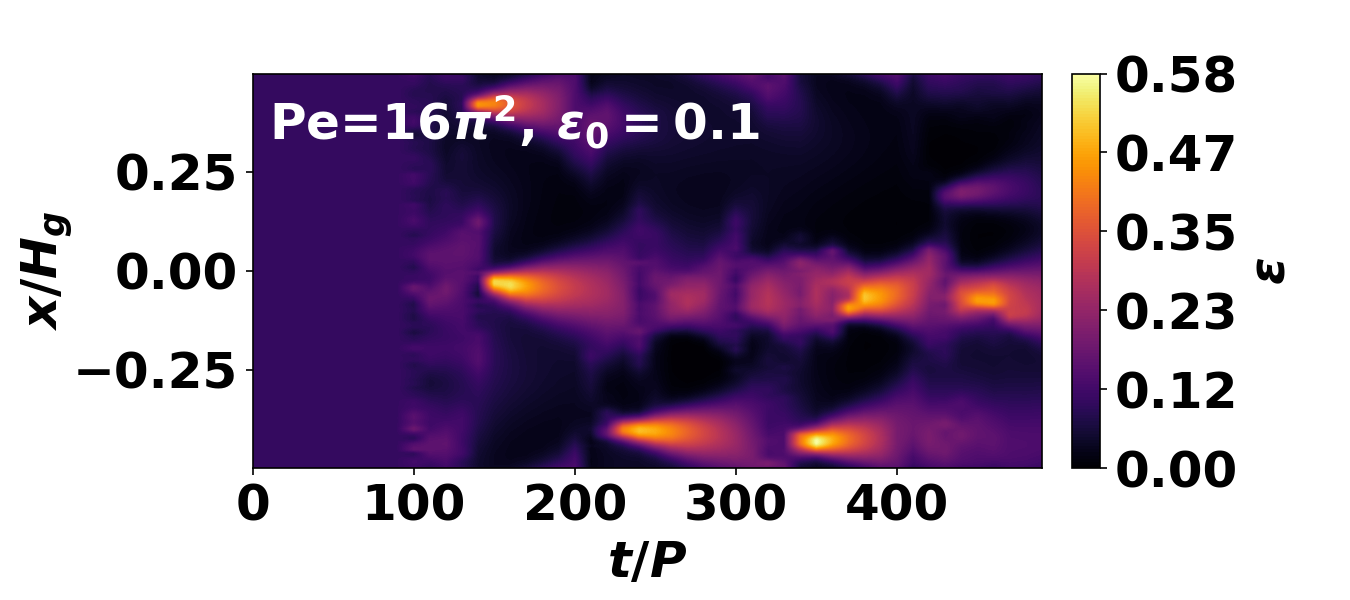}\\
    \caption{Space-time evolution of the dust-to-gas ratios for the COS under stronger thermal diffusion $\peclet=16\pi^2$ for $\epsilon_0=0.01$ (top) and $\epsilon_0=0.1$ (bottom). 
    }
    \label{COS_spacetime_eps_Pe160}
\end{figure}

Fig. \ref{COS_spacetime_eps_Pe160} shows the space-time evolution of the dust-to-gas ratios. Due to the intermittency of zonal flows, dust rings are also transient features lasting only a few tens of orbits, although higher dust-loading ($\epsilon_0=0.1$) slightly extends their lifetimes. Notice, as for the higher $\peclet$ runs above, that $\epsilon$ reaches at most $\sim 0.6$ in the dust rings before its dispersal. 

\section{Parameter study}\label{survey}

We conduct a parameter survey across $\epsilon_0$, $\st$, and $\Pi$. In each set, we vary one of these while keeping the other two fixed to their fiducial values, which are $(\epsilon_0, \st, \Pi)=(0.01, 0.1, 0)$. We return to the reference value of $\peclet = 160\pi^2$. 
We lower the resolution to $N_x\times N_z = 1024\times 512$ to make these surveys computationally feasible. This is expected to give similar results to that at the full resolution; see Appendix \ref{resolution}. 

\subsection{Varying $\epsilon_0$}

\begin{figure}
    \centering
    \includegraphics[width=\linewidth]{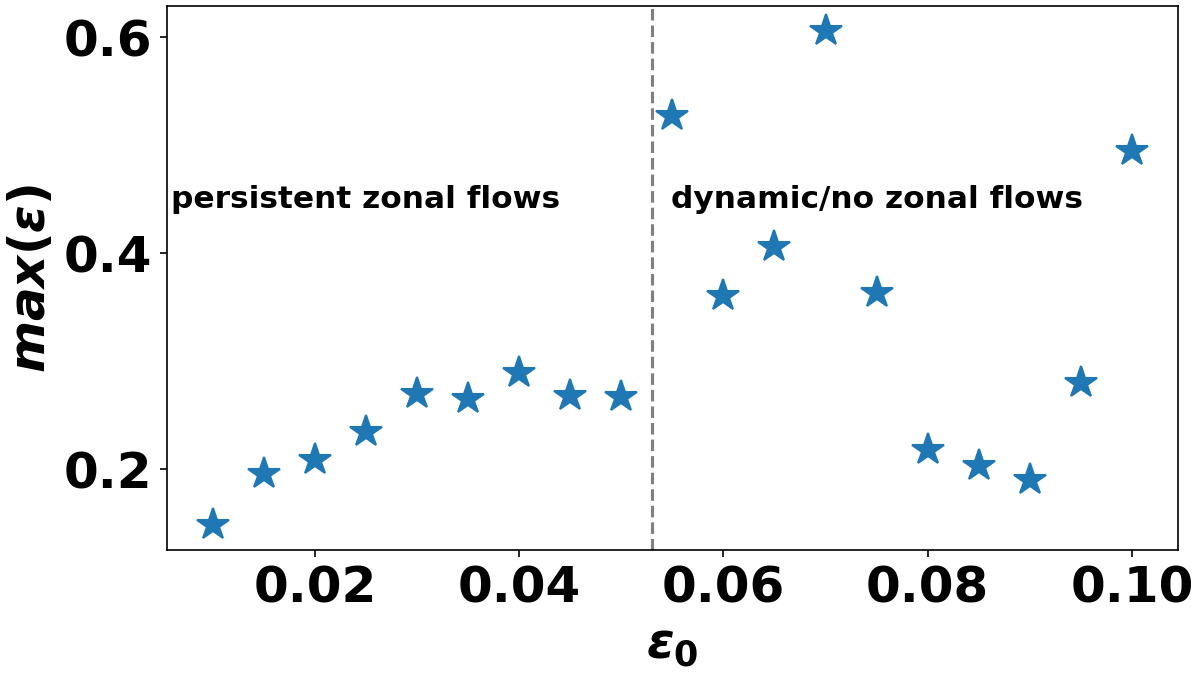}
    \caption{Maximum dust-to-gas ratios averaged between $t\in[500,1000]P$ as a function of the initial dust-to-gas ratio.}
    \label{survey_eps0_epsilon}
\end{figure}

Fig. \ref{survey_eps0_epsilon} shows the maximum dust-to-gas ratios for varying initial values, averaged between $500$ to $1000$ orbits. For $\epsilon_0\lesssim 0.05$, the maximum attainable (time-averaged) $\epsilon \sim 0.3$, regardless of the initial value. In this regime, well-defined zonal flows reach a quasi-steady state. However, for larger $\epsilon_0$, we find no apparent pattern, as in some cases, zonal flows are weak (e.g., $\epsilon_0=0.09$), and there is little concentration. At the same time, some instances exhibit dynamic zonal flows with SI-like instabilities that concentrate dust more appreciably (e.g., $\epsilon_0=0.07$) but these events are transient. Even in those cases, however, $\operatorname{max}(\epsilon) < 1$, which reflects the difficulty of trapping dust in turbulent zonal flows. 

\begin{figure}
    \centering
    \includegraphics[width=\linewidth]{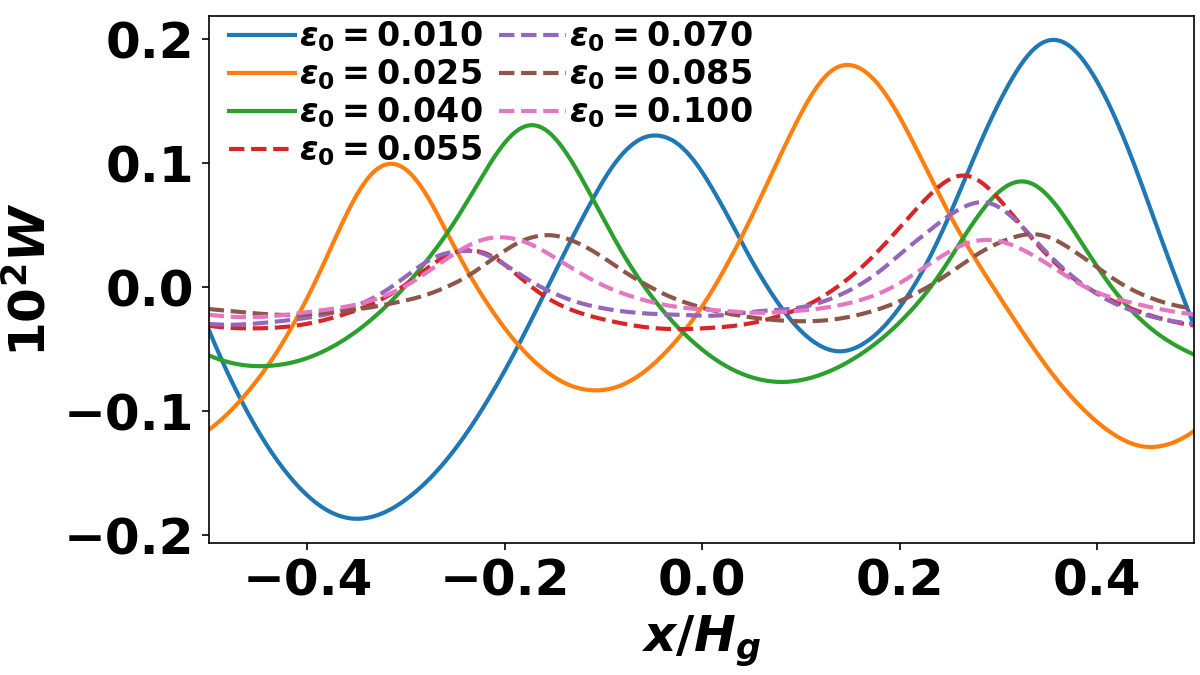}
    \caption{Vertically and time-averaged (from 500 to 1000 orbits) radial profiles of the pressure distribution from the lower-resolution parameter survey across the initial dust-to-gas ratio ($\epsilon_0$).}
    \label{survey_W_profiles}
\end{figure}

In Fig. \ref{survey_W_profiles}, we compare the vertical and time-averaged pressure profiles for selected cases across $\epsilon_0$. While pressure bumps appear in all runs, there is a dichotomy between $\epsilon_0\lesssim 0.05$ and $\epsilon_0\gtrsim 0.05$, with the former having noticeably larger amplitudes (i.e., persistent zonal flows) than the latter. However, the amplitudes within each $\epsilon_0$ regime are similar.

\begin{figure}
    \centering
    \includegraphics[width=\linewidth]{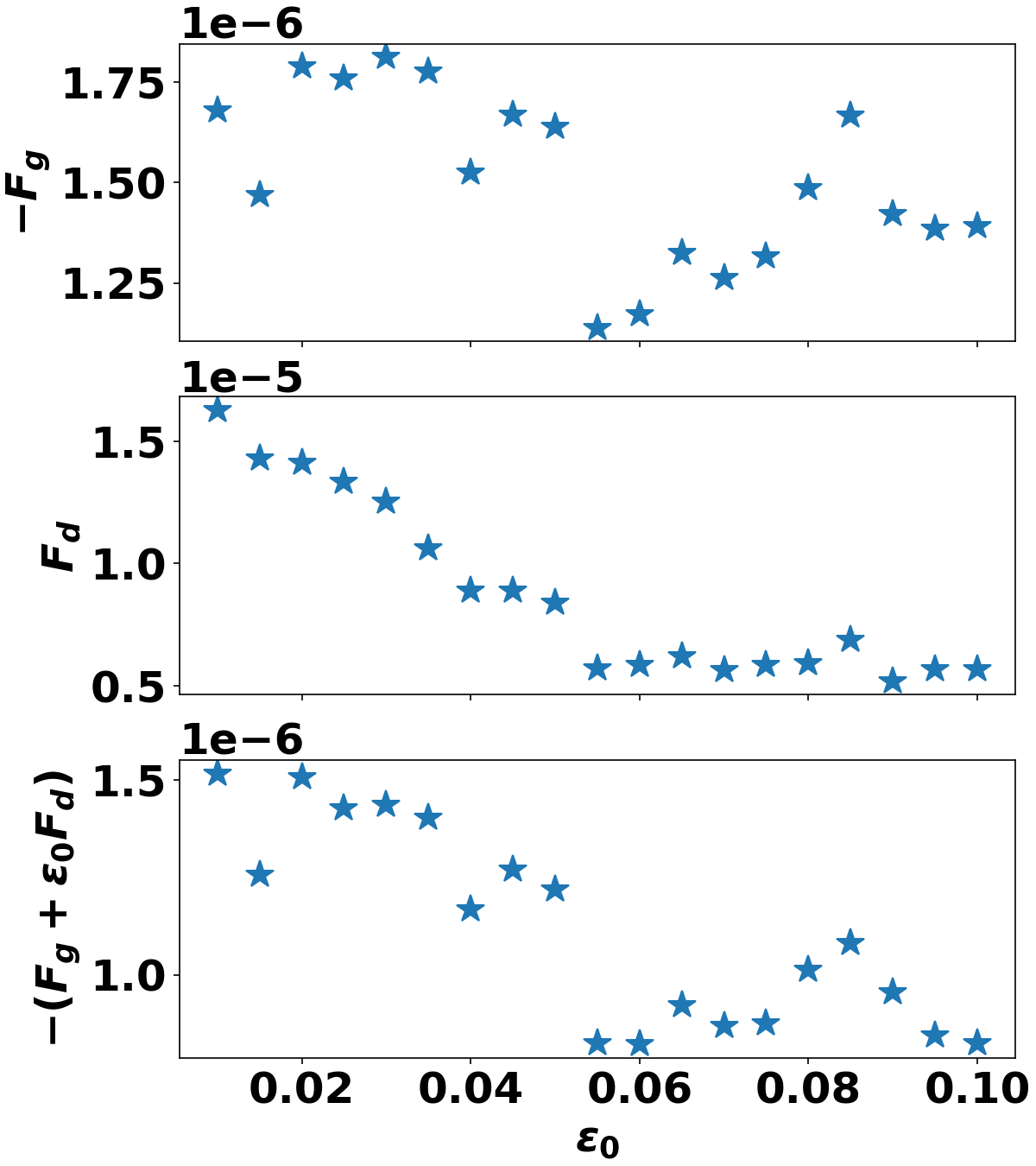}
    \caption{Angular momentum fluxes as a function of initial dust-to-gas ratios. Top: negative of the gas flux; middle: dust flux; bottom: negative of the total flux. Fluxes are averaged over $[500,1000]$ orbits.
    }
    \label{compareAMF_epsilon}
\end{figure}

Fig. \ref{compareAMF_epsilon} shows the time and box-averaged AMFs associated with the gas, dust, and total flux. Fluxes decrease in magnitude with increasing $\epsilon_0$. However, there is considerable scatter in $\Fg$ (with a maximum difference of $\sim 30\%$); while $\Fd$ drops by $\sim 67\%$ by $\epsilon_0=0.07$. This indicates the underlying gaseous COS is not strongly affected by dust feedback for $\epsilon_0\lesssim 0.1$. 

On the other hand, the dust's response weakens with feedback as increasing $\epsilon_0$ reduces drift speeds (see \S\ref{discuss_AMF}). 
Notice that the dust AMFs are almost an order of magnitude larger than those in gas. This is consistent with linear fluxes associated with low-frequency COS modes (see Appendix \ref{linearAMF_low_freq}). $|\Fd|$ decreases with $\epsilon_0$ and saturates for $\epsilon_0\gtrsim 0.07$, meaning dust contributes a more significant mass fraction to the total flux as $\epsilon_0$ further increases.  
Thus, overall, the total AMF becomes less negative with increasing dust abundance. 

\subsection{Varying $\st$}\label{survey_st}

\begin{figure}
    \centering
    \includegraphics[width=\linewidth]{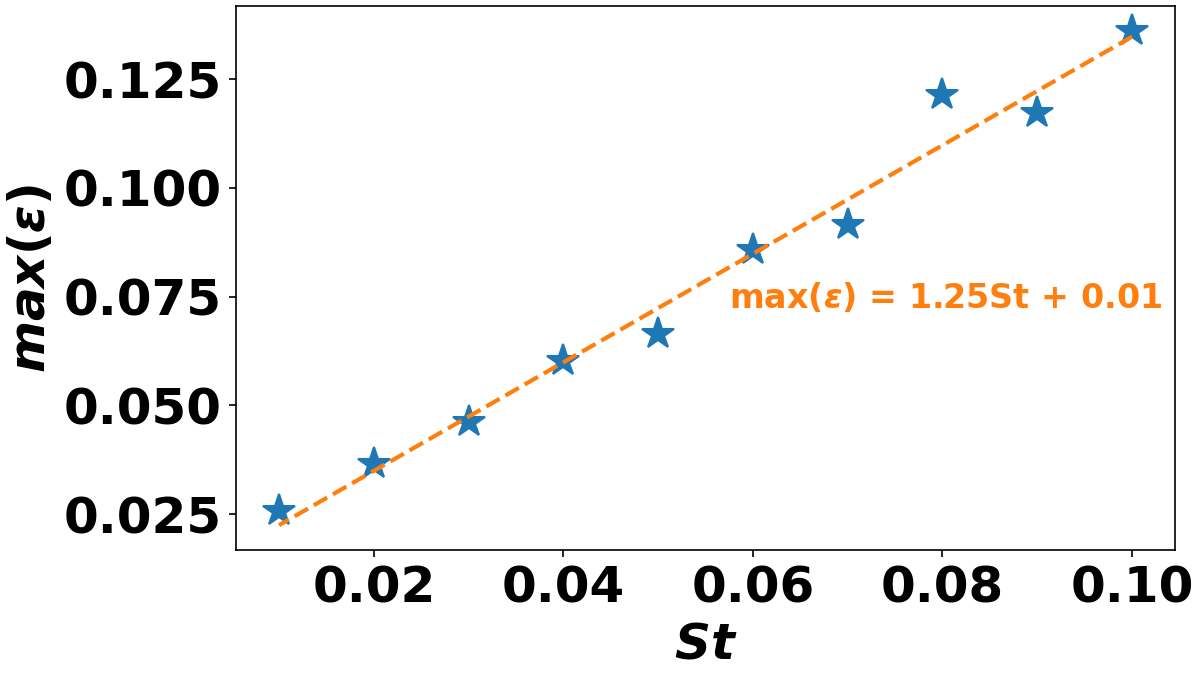}
    \caption{Maximum dust-to-gas ratios averaged between $t\in[500,2000]P$ of simulations with varying Stokes numbers.
    }
    \label{compare_epsilon_st}
\end{figure}

We next vary the degree of dust-gas coupling via $\st$. We extend these simulations to $2000$ orbits to ensure that dust rings reach a quasi-steady state. Fig. \ref{compare_epsilon_st} shows the maximum $\epsilon$, averaged between $500$ and $2000$ orbits. As expected, dust concentrations increase with $\st$. The increase is well-approximated as linear, with $\operatorname{max}(\epsilon)\simeq 1.25\st + 0.01$. For this fit, we impose $\operatorname{max}(\epsilon)\to 0.01$ as $\st\to 0$, since in the perfectly coupled limit $\epsilon$ remains constant. However, this dependence is steeper than one expects from simple diffusion theory. We discuss this in \S\ref{discussion_grain_size}.

\subsection{Varying $\Pi$}

\begin{figure}
    \centering
    \includegraphics[width=\linewidth]{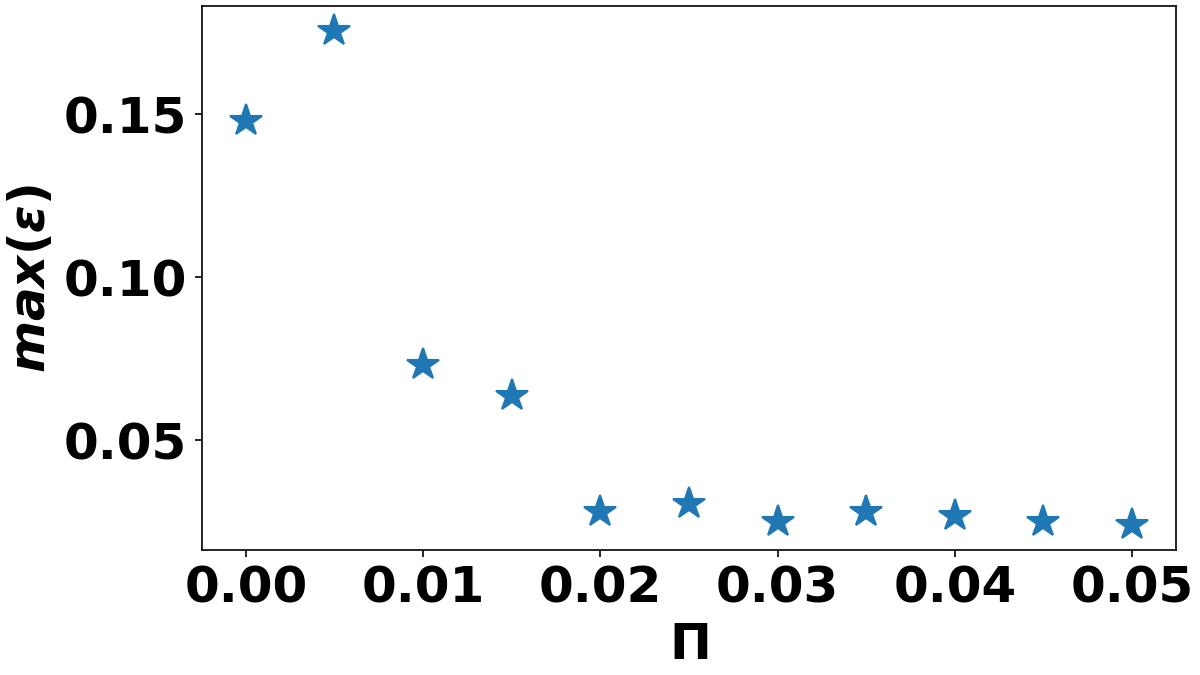}
    \caption{Maximum dust-to-gas ratios averaged between $t\in[500,1000]P$ as a function of the background radial pressure gradient. 
    }
    \label{compare_epsilon_eta}
\end{figure}

Fig. \ref{compare_epsilon_eta} show the maximum dust-to-gas ratios for simulations with $\Pi\in[0,0.05]$. This range is motivated by the fact that $\Pi$ is $O(\hgas)$ in a global disk, and typically $\hgas\simeq 0.05$---$0.1$. Dust concentrations decline rapidly with increasing $\Pi$, or equivalently the background dust drift. As we discuss in \S\ref{dust_concentration_Pi}, this is due to the weak pressure perturbations associated with the COS relative to the background gradient. Zonal flows form in all simulations, and AMFs are similar (not shown). This is expected because the main effect of dust drift at low abundances is to drive a constant background gas flow, which has little impact on the gas evolution. 

\section{Discussion}\label{discussion}

To facilitate the following discussion, it is helpful to summarize the mechanism for zonal flow formation by the pure gas COS as proposed by \citetalias{teed21}. Their interpretation considers how the COS turbulence affects the disk's stability properties. This is measured by the local Rayleigh number $\nrsq_\mathrm{loc} = \left(N_r^2/2\Omega\right)\left(\p_x \theta_x/\p_x j\right)$, representing the competition between a destabilizing radial entropy gradient ($\p_x\theta_x$, where $\theta_x = \theta - x$) and a stabilizing radial angular momentum gradient ($\p_x j$, where $j=\Omega x/2 + \vgy$). The background angular momentum gradient is positive, equal to $\Omega/2$. It is assumed that COS-turbulence drives a negative AMF, $\Fg<0$, whose strength increases with $\nrsq_\mathrm{loc}$, i.e., $\p|\Fg|/\p\nrsq_\mathrm{loc}>0$. 

Next, consider a wave-like perturbation to $j$, say $\dd j $, with a $\dd j > 0$ in $x<0$ and vice versa. The region surrounding $x=0$ has $\p_x \dd j < 0$, offsetting the background positive gradient. This leads to a smaller $\left|\p_x j\right|$ and a larger $\left|\nrsq_\mathrm{loc}\right|$, which increases an inwards flux, $\dd\Fg < 0$. However, this flux is directed towards $x<0$, which already has $\dd j > 0$. This enhances the perturbation there, and the process runs away. The result is layer formation. See \citetalias{teed21} for a detailed description  (their Section 6.2 and Fig. 19). 

\subsection{AMF of dusty gas under geostrophic balance}\label{discuss_AMF}

\citetalias{teed21}'s mechanism relies on a negative AMF. We expect the same requirement for dusty gas but for the total gas-plus-dust AMF, $F<0$. Here, we demonstrate that the specific AMF for dust, $\Fd$, is more positive than the gas, which allows the possibility of $\Fd>0$, which can then reduce $|F|$. This would then work against zonal flow formation. For this discussion, we neglect viscosity and temporarily restore the subscript `0' to denote equilibrium values. 

We begin with the definition 
\begin{align} \label{TVA_vdxprime}
    \dd \vdx = \dd \vgx + \left(\Delta v_x - \Delta v_{x0}\right),
\end{align} 
where $\Delta v_{x0}$ is the equilibrium relative radial drift. For $\st\ll 1$, we can apply the TVA to first order in $\st$, which implies $\dd \vdy =\dd \vgy$ (see Appendix \ref{TVA}). Multiplying Eq. \ref{TVA_vdxprime} by the azimuthal velocity perturbations, we have
\begin{align}\label{Fdust_TVA}
    F_\mathrm{d} = F_\mathrm{g} + \dd\vgy\left(\Delta v_x - \Delta v_{x0}\right).
\end{align}
We expect the second term on the RHS to be positive because dust tends to drift toward pressure maxima. Consider, for simplicity, the case without a background radial pressure gradient ($\eta=0$) so that the equilibrium velocities vanish. Then $\dd \vgy < 0$ corresponds to sub-Keplerian flow, for which dust drifts inwards ($\Delta v_x <0$), and vice versa. Thus the product $\dd\vgy\Delta v_x>0$. 

We can quantify the above argument as follows. We invoke the TVA for $\Delta v_x$,   
\begin{align}\label{TVA_x}
    \Delta v_x = \taus \fgas \left(\p_x W - 2\eta r \Omega^2 + N_r^2\theta\right)
\end{align}
to $O(\st)$, see Eq. \ref{TVA_expand}---\ref{improved_TVA_0}. Here, the gas fraction is $\fgas=1/(1+\epsilon)$. We also define the dust fraction, $\fdust = 1-\fgas$. 
Then, using $\Delta v_{x0} = - 2 \taus \eta r \Omega^2 f_{\mathrm{g}0}$ in the TVA (see Eq. \ref{TVA_Dvx}), we find
\begin{align}\label{Dvx_prime_TVA}
    \Delta v_x - \Delta v_{x0} = \taus \fgas  \left(\p_x W + N_r^2\theta \right) - 2\taus \eta r \Omega^2 \dd \fgas. 
\end{align}

We next assume the gas attains geostrophic balance such that 
\begin{align}
    0 = 2\Omega\vgy - \p_x W - N_r^2\theta + \frac{\epsilon}{\taus}\Delta v_x. \label{geostrophic}
\end{align}
This is motivated by zonal flows corresponding to the conventional geostrophic balance between Coriolis forces and the pressure gradient \citepalias{teed21}. Using the TVA again (Eq. \ref{TVA_x}) to eliminate $\Delta v_x$, we find:
\begin{align}\label{vgy_geostrophic}
    \vgy = \eta r \Omega \fdust + \frac{\fgas}{2\Omega} \left(\p_x W + N_r^2\theta \right).
\end{align}
Subtracting the TVA equilibrium azimuthal gas velocity $v_{\mathrm{g}y0} = f_{\mathrm{d}0} \eta r \Omega$ (see Eq. \ref{TVA_vgy}), we find 
\begin{align}\label{TVA_pressure_pert}
     \fgas \left(\p_x W + N_r^2\theta \right) = 2\Omega \dd \vgy - 2 \eta r \Omega^2 \dd \fdust.
\end{align}

Finally, combining Eqs. \ref{Fdust_TVA}, \ref{Dvx_prime_TVA}, and \ref{TVA_pressure_pert} to eliminate the pressure and buoyancy variables give 
\begin{align}
    F_\mathrm{d} = F_\mathrm{g} + \dd \vgy \left[\taus\left(2\Omega \dd \vgy - 2 \eta r \Omega^2\dd \fdust\right) - 2 \taus \eta r \Omega^2 \dd \fgas\right]\notag,
\end{align}
{ which yields}
\begin{align}
    F_\mathrm{d} = F_\mathrm{g}  + 2 \st \left(\dd \vgy\right)^2, \label{Fd_model}
\end{align}
where we used $\fdust + \fgas = 1$ and thus $\dd \fdust + \dd \fgas = 0$. Since the second term is positive definite, the dust AMF is always more positive  than the gas AMF. We usually find $F_\mathrm{d}>0$, implying the second term outweighs the first in practice.

\begin{figure}
    \centering
    \includegraphics[width=\linewidth]{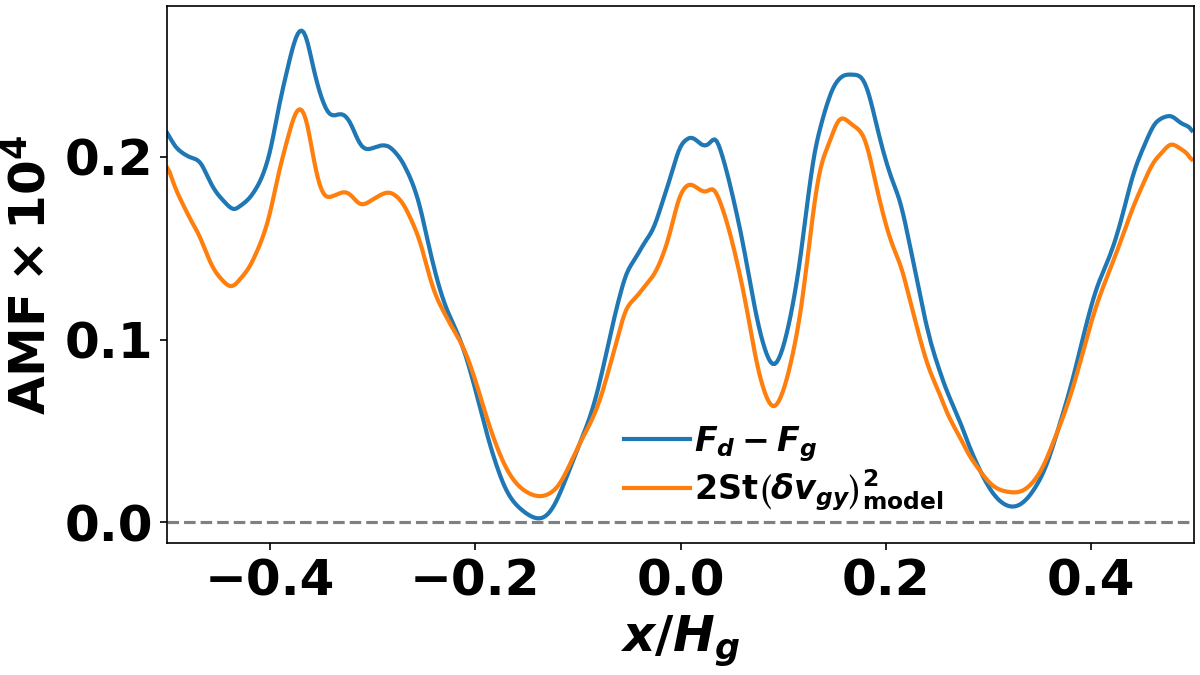}
    \caption{{ Vertically and time-averaged (between $t\in[500,1000]P$) angular momentum flux difference associated with dust and gas velocity fluctuations (blue), and its value according to the model given by Eq. \ref{Fd_model} with $\dd \vgy$ given via Eq. \ref{vgy_geostrophic} (orange). The close match shows that the flux difference results from geostrophic balance in the gas and dust-trapping in response to pressure fluctuations and buoyancy. 
    }}
    \label{FdFg}
\end{figure}

In Fig. \ref{FdFg}, we plot the vertically and time-averaged $\Fd-\Fg$ as measured from our fiducial run (blue) and compare with that according to Eq. \ref{Fd_model} { (orange), with $\delta\vgy$ given via Eq.  \ref{TVA_pressure_pert}, which encapsulates geostrophic balance and the TVA. The average dust AMF is indeed more positive than the gas AMF and is consistent with the model. 
}

Eq. \ref{Fd_model} relates $\Fd$ and $\Fg$ but cannot be used to predict their relative magnitudes. Instead, this can be done within linear theory. In Appendix \ref{linear_dusty_AMF}, we show that low-frequency COS modes induce a positive dusty AMF for sufficiently large Stokes numbers relative to the COS growth rates. Below, we also use it to estimate the dust-to-gas ratios beyond which the dust AMF overcomes the gas AMF and find consistency with simulation results. 


\subsection{Limitations to dust concentration by COS}

\subsubsection{By dust feedback}

Our simulations show that dust feedback mitigates zonal flow formation by the COS and limits dust concentration in existing zonal flows. At first, it is somewhat surprising that this manifests at $\epsilon\sim 0.1$, where feedback is usually considered negligible. However, this is consistent with the argument that the effect stems from the positive dust AMF ($\Fd>0$) counteracting the negative gas AMF ($\Fg<0$) that necessitates zonal flow formation. The fact that dust drift is sensitive to pressure fluctuations means that the dust AMF can overcome the gas AMF, even when $\epsilon \lesssim 1$. 

We can estimate the critical $\epsilon$ by considering linear COS modes as discussed in Appendix \ref{linear_dusty_AMF}. For slowly-growing, low-frequency modes, the dust-weighted AMF ratio $|\epsilon \Fd/\Fg|\sim \epsilon \st/s$ for $\st \gg s$, where $s$ is the modes'  dimensionless growth rate. (Note that $\st > s$ is necessary for $\Fd>0$.) For the inviscid COS, $s = \nrsq/4 $ \citep{latter16,lehmann23}. We then find the weighted dust-to-gas AMF ratio exceeds unity when $\epsilon \gtrsim \nrsq/(4\st)$. This evaluates to  $\epsilon = 0.25$ for our fiducial parameters. Indeed, the maximum dust-to-gas ratios first saturate at this value in Fig. \ref{survey_eps0_epsilon}. 

In a global disk, we expect $\left|\nrsq\right|\sim \hgas^2$ on dimensional grounds. Then, setting $s\sim \hgas^2$ and the AMF ratio to unity, we find the critical dust-to-gas ratio 
\begin{align}
\epsilon \gtrsim \frac{\hgas^2}{\st} \quad \text{to inhibit zonal flows,}
\end{align}
which can be less than unity. For example, with $\hgas\simeq 0.1$ and grain sizes $\st\simeq 0.1$, we have $\epsilon\gtrsim 0.1$. For $\hgas\simeq 0.05$ and smaller grains $\st\simeq 0.01$, the critical $\epsilon\simeq 0.25$. 

\subsubsection{By grain size}\label{discussion_grain_size}
In \S\ref{survey_st}, we observed that the maximum dust-to-gas ratio attained increases linearly with $\st$. This differs from the conventional advection-diffusion description of dust trapping with a constant diffusion coefficient \citep[e.g.][]{dullemond18}. Here, one expects a Gaussian zonal flow with width $\Delta_\mathrm{ZF}$ to produce a Gaussian dust ring with width $\Delta_\mathrm{d} \simeq \sqrt{\alpha_\mathrm{d}/\st}\Delta_\mathrm{ZF}$, where $\alpha_\mathrm{d}$ is a dimensionless radial diffusion coefficient for the dust, assumed to be $\ll \st$, which results from turbulent stirring by the gas. Given a total dust mass, we expect $\operatorname{max}(\epsilon)\propto \left(\st/\alpha_\mathrm{d}\right)^{1/2}/\Delta_\mathrm{ZF}$. Thus, if $\alpha_\mathrm{d}$ and $\Delta_\mathrm{ZF}$ are independent of $\st$, then $\operatorname{max}(\epsilon)\propto \sqrt{\st}$. 

Our empirical result $\operatorname{max}(\epsilon)\sim \st$ can be recovered if $\alpha_\mathrm{d}\sim \st^{-1}$ while $\Delta_\mathrm{ZF}$ remains constant. This should be tested in COS simulations with Lagrangian particles to measure dust diffusion coefficients explicitly. 



\subsubsection{By a background drift}\label{dust_concentration_Pi}

Our simulations with $\Pi\neq 0$ demonstrate the resulting background dust drift significantly mitigates their concentrations in zonal flows. This can be understood, again, within the TVA as a competition between the local and global pressure gradients. Neglecting buoyancy, Eq. \ref{improved_TVA_0} shows that these become comparable when $
\p_x W \sim 2 \eta r \Omega^2$. Again considering a zonal flow width $\Delta_\mathrm{ZF}$, we can estimate the critical $\Pi$ as
\begin{align}
    \Pi \gtrsim \left(\frac{W}{c_s^2}\right)\left(\frac{\Hgas}{\Delta_\mathrm{ZF}}\right) \quad \text{to inhibit dust trapping.}
\end{align}
In our fiducial run, zonal flows have $\operatorname{max}(W)\sim 0.003c_s^2$ and $\Delta_\mathrm{ZF}\sim 0.1\Hgas$ (Fig. \ref{COS_spacetime_W_Pe1600}, \S\ref{result_fid}), which imply even $\Pi\sim 0.03$ will interfere will dust trapping. Indeed, Fig. \ref{compare_epsilon_eta} shows that dust concentrations become negligible for $\Pi\gtrsim 0.02$. 

\subsection{Implications for planetesimal formation}

Our simulations suggest feedback sets an upper limit to dust concentration by the COS zonal flows with $\epsilon\lesssim O(10^{-1})$, even for relatively large grains with $\st=0.1$. This is far smaller than that needed for direct gravitational collapse \citep{shi13} or the SI, requiring $\epsilon\gtrsim 1$ to drive meaningful dust clumping. 


Furthermore, zonal flows' weak pressure perturbations make dust concentrations vulnerable to background dust drift. Therefore, a small global radial pressure gradient is desirable. Unfortunately, this also reduces the magnitude of the buoyancy frequency needed to drive COS in the first place. For definiteness, consider local power-law density and temperature profiles with $\rhog\propto r^{-\mu}$ and $T\propto r^{-q}$, respectively. The buoyancy parameter becomes
\begin{align*}
    \nrsq = \frac{4\Pi^2}{\gamma}\left(1 - \frac{\gamma \hgas \mu}{2\Pi}\right),  
\end{align*}
\citep[][see their Eqs. 15---16]{lehmann24}. For instability, it is necessary to have $\nrsq>0$, which implies a { shallow} or rising density profile ($\mu\lesssim 0$), assuming $\Pi>0$. Requiring $\nrsq$ to be sufficiently large for significant growth \emph{and} $\Pi$ to be sufficiently small to produce effective dust traps requires a sharply increasing density profile, perhaps unrealistically so. 


One way to overcome the above difficulties is dust trapping by COS-induced vortices. This is beyond the scope of our axisymmetric models. However, \cite{raettig15, raettig21} have shown that such vortices concentrate dust grains effectively, provided they form in the first place. Indeed, their simulations are first evolved without dust to allow the pure gas COS to create a large-scale vortex. This process likely involves the break-up of COS-induced zonal flows (\citealt{latter16}; \citetalias{teed21}). If zonal flows are a prerequisite to vortex formation, our results suggest that the disk cannot be too dusty initially (i.e., $\epsilon\lesssim 0.1$). We speculate that COS-assisted planetesimal formation should be more relevant to dust-poor disk regions. 

\subsection{Caveats and outlook}

Our models adopt the minimal geometry for the COS: axisymmetric and unstratified. These simplifications need to be relaxed in the future. 

Equating our horizontal box sizes to $\sim \Hgas$ in a corresponding global disk, zonal flows in our simulations are narrow, with widths $\sim 0.1\Hgas$ and even smaller in the dust. Such structures are likely unstable to the Rossby Wave Instability \citep[RWI,][]{lovelace99,li00} and its dusty analogs \citep{liu23}, leading to vortex formation \citep{li01}.  
On the other hand, we find COS zonal flows are dynamic and internally turbulent. It is unclear how vortex formation by RWI-like processes would be affected, particularly for dusty zonal flows. This will need to be studied using models including the azimuthal direction. 

Our unstratified disk models, with $|z|\leq 0.25\Hgas$, apply to the gas since a stratified gas disk has a Gaussian distribution with scale height $\Hgas$. Thus, the gas density drop over our vertical domain is negligible. On the other hand, in a weakly turbulent disk, the expected dust scale height is given by
\begin{align}
    \frac{\Hdust}{\Hgas}=\sqrt{\frac{M_z^2\tau_\mathrm{eddy}}{\st}},
\end{align}
where $M_z$ is the average vertical Mach number and $\tau_\mathrm{eddy}$ is the eddy turnover timescale normalized by $\Omega^{-1}$ \citep[e.g.][]{lin19}. For $\st=0.1$, $M_z\sim 0.02$ as measured our fiducial run, and taking $\tau_\mathrm{eddy}\lesssim 1$ gives $\Hdust\lesssim 0.06\Hgas$. That is, the dust should be stratified. Future simulations should thus include vertical gravity, at least on the dust (M. Lehmann \& M.-K. Lin, in preparation). 

However, stratification adds significant complexity as it enables a plethora of other instabilities. This includes the gaseous VSI and generalizations of the COS \citep{lehmann23, klahr23, klahr24}, as well as drag instabilities such as the Dust Settling Instability \citep{krapp20} and the Vertically Shearing Streaming Instability \citep{lin21}. As preparation for such a study, it is perhaps worth first investigating the impact of vertical domain size and, particularly, boundary conditions on the COS within the unstratified setting.    

{ 
Finally, the Boussinesq approximation requires that gas velocities remain subsonic and density perturbations remain small. This is expected for COS-driven turbulence since they comprise inertial waves, which are locally incompressible \citep{balbus03}. We thus do not expect dust dynamics (which responds primarily to pressure variations) to be significantly affected on small scales. 

On the other hand, simulating the COS across a more significant portion of the global disk, or in 3D where vortex formation occurs and would launch spiral density waves, requires a compressible treatment as these are expected to significantly modify the underlying disk structure \citep{lehmann24}.  
}

\section{Summary}\label{summary}

In this paper, we conduct high-resolution spectral simulations of the COS in a dusty PPD. We adopt the conventional Boussinesq shearing box framework for studying convection in a local disk patch. We add dust as a second, pressureless fluid coupled to the gas via drag forces. Our fiducial setup considers a relatively steep entropy gradient with a squared buoyancy frequency $N_r^2 = -0.1\Omega^2$, weak thermal diffusion with Peclet number $\peclet\simeq1600$, and large grains with Stokes number $\st=0.1$ and a dust-to-gas mass ratio $\epsilon=0.01$. We fix the Reynolds number to $\reynolds=10^5$, { equivalent to an $\alpha$ viscosity of $10^{-5}$ in our setup}, which sets the gas viscosity and dust diffusion coefficients. 

In the limit of negligible dust feedback, our setup produces quasi-steady zonal flows (pressure bumps) that concentrate dust to $\sim 10$ times its initial density. The internal turbulence of zonal flows limits concentration. Concentration factors typically decrease when feedback is included or upon increasing the initial $\epsilon$. We find maximum dust-to-gas ratios  $\sim 0.2$, after which zonal flows weaken. When initialized with $\epsilon=0.1$, we find zonal flows are suppressed. 
We interpret these as a result of the positive dust AMF offsetting the negative gas AMF needed to form zonal flows. 

We also find that a background dust drift, usually attributed to a global radial pressure gradient, strongly reduces the dust-trapping capability of COS-produced zonal flows. This is due to the weak pressure perturbations associated with zonal flows compared to typical values of the global radial pressure gradients. However, this gradient cannot be too small as it also sets the disk's buoyancy response, which is responsible for driving the COS in the first place. 

We conclude that COS-driven zonal flows are not directly conducive to triggering planetesimal formation. However, relaxing the unstratified and axisymmetric approximations, thereby allowing dust settling and vortex formation, respectively, will be necessary to assess the impact of the COS in a realistic disk. 


\begin{acknowledgments}
{We thank the anonymous referee for a helpful report.} This work is supported by the National Science and Technology Council (grants 112-2112-M-001-064-, 113-2112-M-001-036-, 113-2124-M-002-003-) and an Academia Sinica Career Development Award (AS-CDA-110-M06). Simulations were performed on the Kawas cluster at ASIAA, the Academia Sinica Grid Computing clusters, and the Taiwania-3 cluster
at the National Center for High-performance Computing (NCHC). We thank NCHC for providing computational and storage resources. 
\end{acknowledgments}

\appendix \label{appendix}

\section{Terminal velocity approximation}\label{TVA}
We can simplify the relative velocity equation (Eq. \ref{gas_based_Dv}) when $\st\ll 1$. In this limit, applicable to dust tightly coupled to the gas, the relative dust-gas drift is expected to be small. We thus neglect terms quadratic in $\Dv$. We further assume that dust reaches terminal velocity following the gas 
\begin{align}
\frac{\p\Dv}{\p t} + \left(\vg\cdot\nabla\right)\Dv = 0.
\end{align} 
We are then left with 
\begin{align}
\left(\Dv\cdot\nabla\right)\vg  =   2\Omega \Dvy \hat{\bm{x}} - \frac{\Omega}{2}\Dvx \hat{\bm{y}} + 
\nud\nabla^2\Dv - \frac{(1+\epsilon)}{\taus}\Dv + \nabla W - 2\eta r \Omega^2\xhat + N_r^2\theta\xhat.\label{improved_tva}
\end{align}
Without dissipation ($\nud=0$), Eq. \ref{improved_tva} is an algebraic equation for the components of $\Dv$ that, in principle, can be solved explicitly. 
Here, we instead assume the expansion 
\begin{align}
    \Dv = \Delta\bmv^{(0)} \st + \Delta\bmv^{(1)} \st^2 + \cdots, \label{TVA_expand}
\end{align}
which vanishes as $\st\to 0$ as expected on physical grounds. Inserting this into Eq. \ref{improved_tva}, we find at zeroth order in $\st$:
\begin{align}
\Delta\bmv^{(0)} = \frac{\fgas}{\Omega} \left(\nabla W - 2\eta r \Omega^2\hat{\bm{x}} + N_r^2\theta\xhat\right), \label{improved_TVA_0}
\end{align}
which corresponds to the classical terminal velocity approximation \citep[e.g.][]{lovascio19} with the addition of radial buoyancy. 

At order $\st$, we find:
\begin{align}
    \Delta\bmv^{(1)} = \frac{\fgas}{\Omega}&\left\{\nud\nabla^2\Dv^{(0)} 
     -\frac{\Omega}{2}\Delta v_x^{(0)}\yhat - \Delta \bm{v}^{(0)}\cdot\nabla\vg\right\},\label{improved_TVA_1}
\end{align}
where we have used the fact that $\Delta v_y^{(0)}=0$ at all times for axisymmetric flow. 


Next, we insert the TVA to simplify the drag term on the RHS of Eq. \ref{gas_based_mom}, so that
\begin{align}
    \frac{\epsilon}{\taus}\Dv \to &\epsilon\Omega\left[\Delta\bmv^{(0)} + \Delta\bmv^{(1)}\st\right]\notag\\ & =\fdust\left(\nabla W - 2\eta r \Omega^2\hat{\bm{x}} + N_r^2\theta\xhat \right) 
    + \st\fdust\left\{\nud \nabla^2\Dv^{(0)}
     -\frac{\Omega}{2}\Delta v_x^{(0)}\yhat - \Delta \bm{v}^{(0)}\cdot\nabla\vg\right\}.
    \label{gas_based_TVA_mom}
\end{align}
The gas-based formulation with the TVA consists of Eqs. \ref{gas_based_gas}---\ref{gas_based_energy},
Eq. \ref{gas_based_dg_Q}, with Eq. \ref{gas_based_TVA_mom} replacing the dust-gas drag term. The evolutionary equation for $\Dv$ (\ref{gas_based_Dv}) is then dropped. 

\subsection{TVA equilibrium}

In the TVA, the equilibrium velocities are given by 
\begin{align}
    &\vgx = \frac{2\st\epsilon}{(1+\epsilon)^2}\eta r \Omega,\\
    &\vgy = \frac{\epsilon}{1+\epsilon}\eta r \Omega, \label{TVA_vgy} \\
    &\Dvx = -\frac{2\st}{1+\epsilon}\eta r \Omega,\label{TVA_Dvx}\\
    &\Dvy = \frac{\st^2}{(1+\epsilon)^2}\eta r \Omega,
\end{align}
while the equilibrium vertical velocities remain zero, $\vgz=\Dvz=0$. The above TVA equilibrium can be obtained from Eqs. \ref{eqm_vgx_full}---\ref{eqm_Dvz_full} by setting $\st^2\to0$ in the denominators. Note that
\begin{align}
    &\Dvx^{(0)} = -2\fgas\eta r \Omega,\\
    &\Dvy^{(1)} = \fgas^2\eta r \Omega,
\end{align}
while $\Dvx^{(1)} = \Dvy^{(0)}=0$.


%

\subsection{Linearized drag forces in the TVA}

Linearizing the drag force under the TVA (Eq. \ref{gas_based_TVA_mom}) yields 

\begin{align}
    &\dd\left(\frac{\epsilon\Dv}{\taus}\right) = \fdust\left( 1 - \frac{\nud k^2 \fgas \st }{\Omega}\right)\left(\ik\dd W + N_r^2\dd\theta\xhat\right) -2\eta r \Omega^2 \fgas^2 \left(1 + \frac{\nud k^2\fdust\st}{\Omega}\right)\dd\epsilon \xhat\notag\\
    &\phantom{\dd\left(\frac{\epsilon\Dv}{\taus}\right) =}
    -\frac{1}{2}\st\fdust\fgas\ikx\dd W\yhat 
    +\st\eta r \Omega^2\fgas^2 \left(\fgas - \fdust\right)\dd\epsilon\yhat
    -\frac{1}{2}\st\fdust\fgas N_r^2\dd\theta \yhat 
    -\ikx \st \fdust \Dvx^{(0)}\dd\vg,
\end{align}
where we used
\begin{align}
    \dd\Dv^{(0)} = 2\eta r \Omega \fgas^2\dd\epsilon\xhat  + \frac{\fgas}{\Omega}\left(\ik\dd W + N_r^2\dd\theta\xhat\right). \label{Deltav0_pert}
\end{align}

Inserting the above into the linearized gas momentum equation (Eq. \ref{linear_vg}), we obtain

\begin{align}
&\sigma \dd\vg + \ikx \vgx \dd\vg = 2\Omega\dd\vgy\xhat 
- \frac{\Omega}{2}\dd\vgx\yhat  - \fgas\left( 1 + \frac{\nud k^2 \fdust \st }{\Omega}\right)\left(\ik\dd W + N_r^2\dd\theta\xhat\right)   -2\eta r \Omega^2 \fgas^2 \left(1 + \frac{\nud k^2\fdust\st}{\Omega}\right)\dd\epsilon\xhat   \notag\\
&\phantom{\sigma \dd\vg + \ikx \vgx \dd\vg =} 
-\frac{1}{2}\st\fdust\fgas\ikx\dd W\yhat  +\st\eta r \Omega^2\fgas^2 \left(\fgas - \fdust\right)\dd\epsilon\yhat -\frac{1}{2}\st\fdust\fgas N_r^2\dd\theta \yhat 
    -\ikx \st \fdust \Dvx^{(0)}\dd\vg
- \nu k^2\dd\vg.\label{linear_vg_TVA}
\end{align}

\subsection{COS and SI in the TVA}\label{COS_SI_TVA}

We demonstrate the utility and limitations of the TVA by recomputing the linear COS and SI modes presented in \S\ref{linear} under the TVA. We solve the same linearized equations as in that section but replace the gas momentum equation with Eq. \ref{linear_vg_TVA} and drop the relative drift equation. 

\begin{figure*}
    \centering
    \includegraphics[scale=0.67,clip=true, trim=0cm 0cm 3cm 0cm]{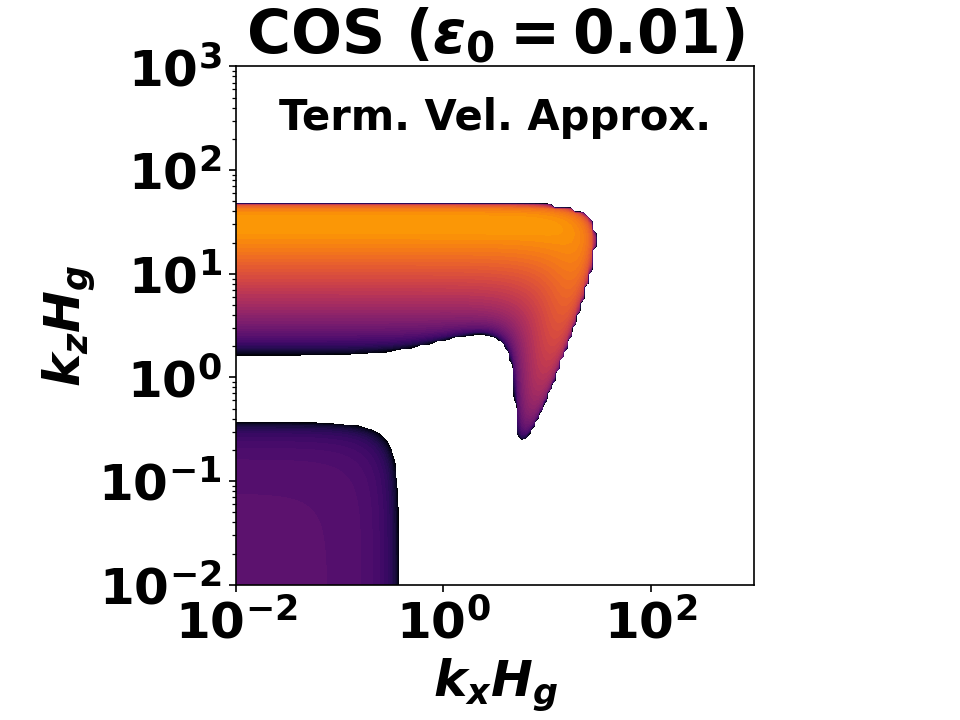}\includegraphics[scale=0.67,clip=true, trim=2.5cm 0cm 0cm 0cm]{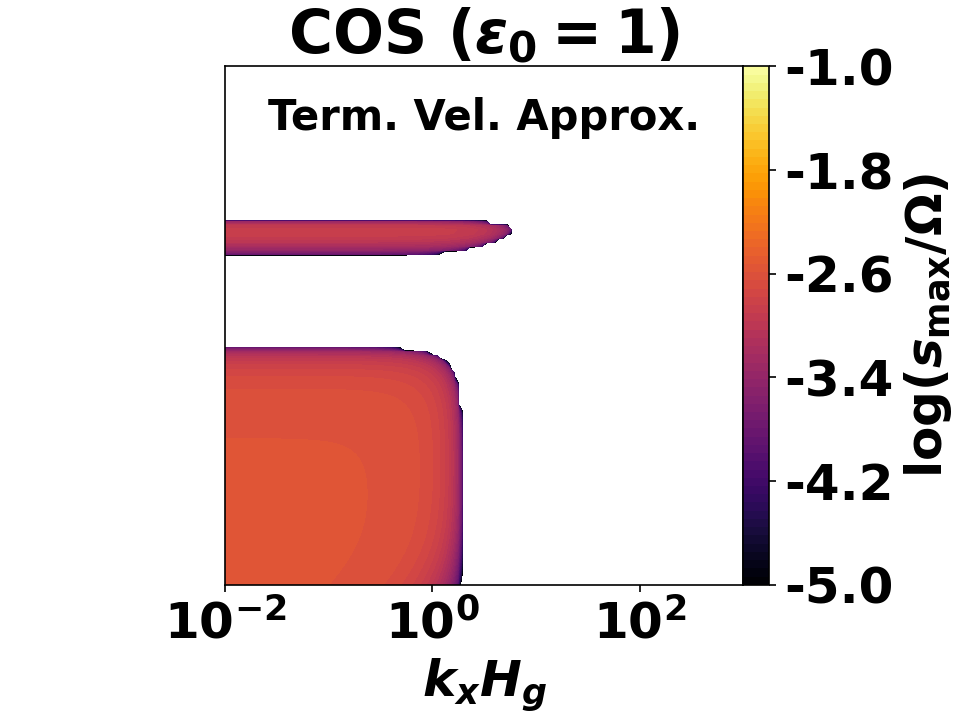}
    \caption{Same as Fig. \ref{COS_exact} but in the terminal velocity approximation.}
    \label{COS_TVA}
\end{figure*}

Fig. \ref{COS_TVA} shows results for the COS and are almost indistinguishable from the full treatment (Fig. \ref{COS_exact}). For $\epsilon_0=1$, growth rates are underestimated in the TVA and modes with $3 \lesssim K_z\lesssim 15$ are entirely missed. This indicates that the TVA is only appropriate for modeling COS at low dust abundances.   


\begin{figure}
    \centering
    \includegraphics[scale=0.67,clip=true, trim=0cm 0cm 0cm 0cm]{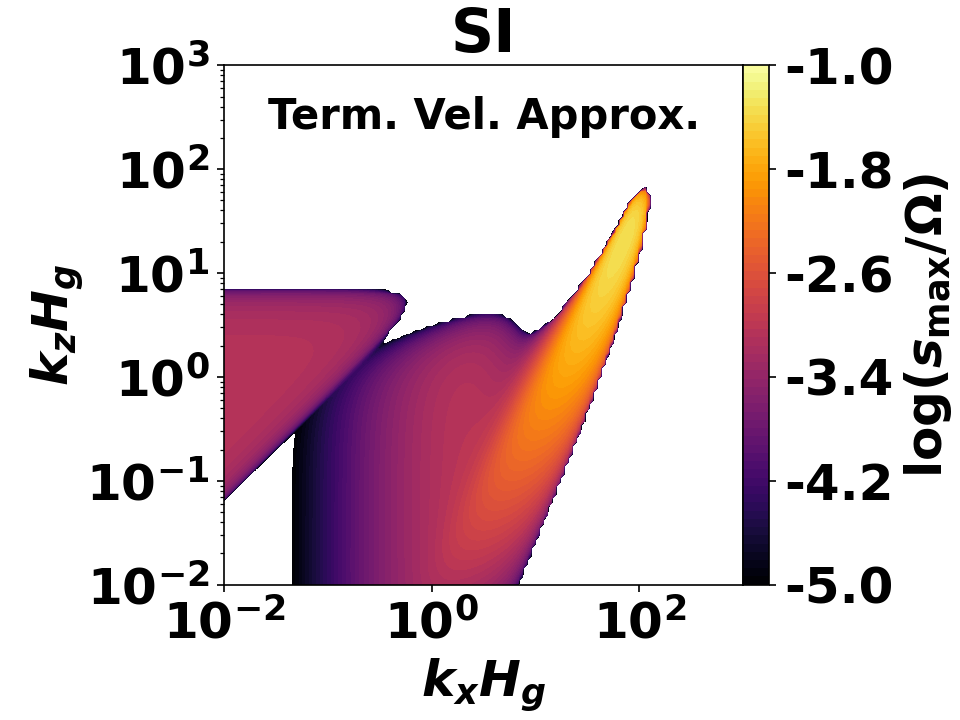}
    \caption{Same as Fig. \ref{SI_exact} but in the terminal velocity approximation.}
    \label{SI_TVA}
\end{figure}

On the other hand, Fig. \ref{SI_TVA} shows results for the SI, and while the most unstable modes are reproduced (cf. Fig. \ref{SI_exact}), spurious modes also appear in the TVA as the `triangle' region to the left, which are absent in the full treatment; see Fig. \ref{SI_exact}. (We find these spurious modes vanish in the isothermal limit  with small $\peclet$.) Although one expects the most unstable modes --- which are correctly captured by the TVA --- to dominate, these spurious modes may pollute simulations over long timescales. Thus, the TVA is not recommended for simulating the SI, unless the spurious modes can be filtered out, e.g. by choosing the appropriate domain size.

\section{Code test}

We test our implementation of the gas-based equations (\ref{gas_based_gas}---\ref{gas_based_energy}, \ref{gas_based_Dv}, and \ref{gas_based_dg_Q}) in the \textsc{dedalus} code by comparing the growth rates of the COS and SI obtained from the simulations and that from linear theory (\S\ref{linear}). We also test an implementation with the TVA, in which case the dust-gas drag term in Eq. \ref{gas_based_mom} is replaced by Eq. \ref{gas_based_TVA_mom}, and the linearized gas momentum equation (\ref{linear_vg}) is replaced by Eq. \ref{linear_vg_TVA}. 

We consider the same disk parameters as that in \S\ref{linear} and Figs. \ref{COS_exact} --- \ref{SI_exact}. The simulations are initialized with a linear eigenmode with wavenumbers $K_{x,z}$ and its amplitude normalized such that the azimuthal velocity perturbation $\delta v_y = 10^{-6}c_s$. We choose the box size to be one wavelength in each direction, i.e. $L_{x,z}=\left(2\pi/K_{x,z}\right)\Hgas$. For the COS, we choose $K_x = 1$ and $K_z=\sqrt{\peclet}\simeq 40$ ($\epsilon_0=0.01$) and $K_z=20$ ($\epsilon_0=1$).  For the SI, we set $K_x = 80$ and $K_z=20$. We use $N_x=N_z=64$ spectral modes in each direction. For the full treatment, we use a dealising factor of $3/2$, while for the TVA runs, we use a factor of $2$ because of the higher degree of nonlinearity in the drag term. We remark that, although there are fewer TVA equations, the simulations did not run faster than the full treatment because of the higher complexity of the TVA equations and the larger dealiasing factor employed. For these tests, we set the maximum time step to $0.1\taus$. 

\begin{figure}
    \centering
    \includegraphics[scale=0.455,clip=true, trim=0cm 0cm 0cm 0cm]{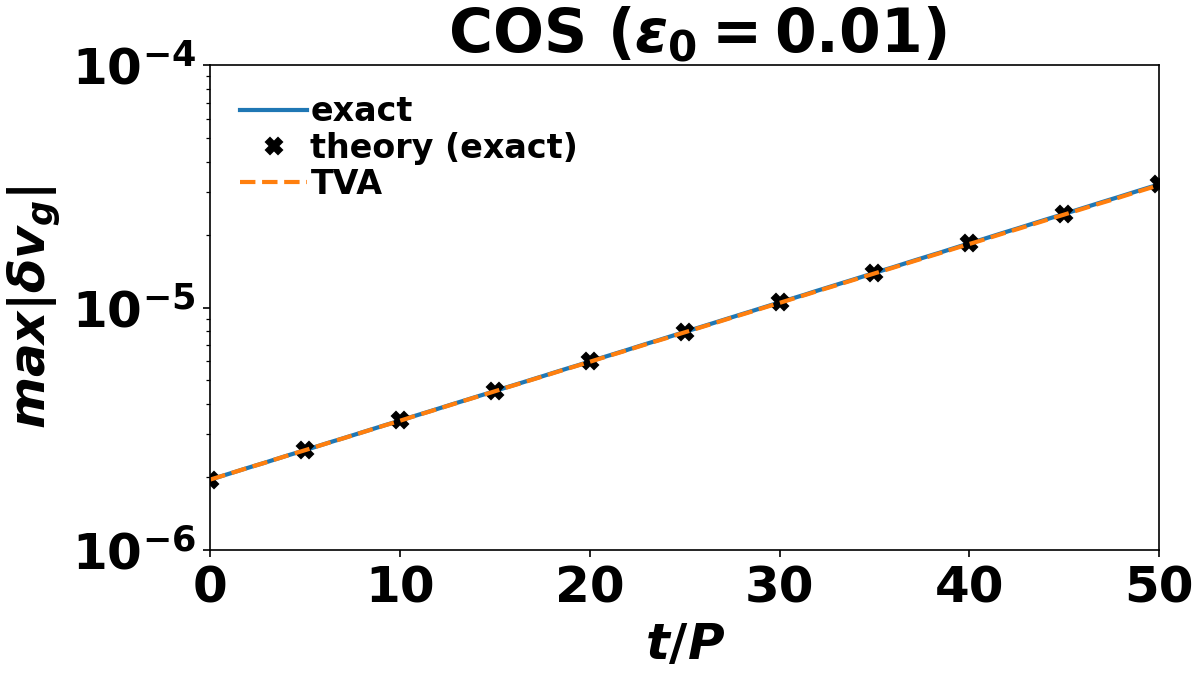}\includegraphics[scale=0.455,clip=true, trim=1cm 0cm 0cm 0cm]{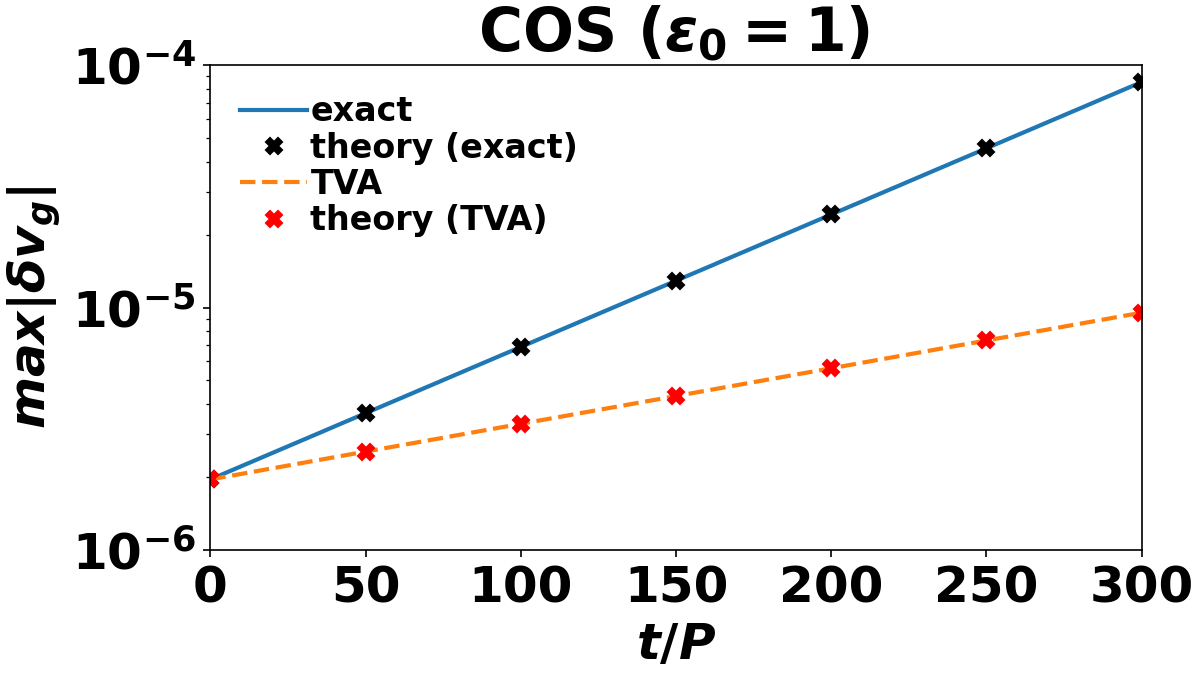}
    \caption{Linear growth of the COS simulated by \textsc{dedalus} (lines), compared with analytic growth rates (asterisks). Dust drag is treated via the full differential velocity equation (blue solid) or the TVA (orange dashed). Left panel: a dust-poor disk with $\epsilon_0=0.01$ (only the exact analytic growth rate is shown since it is similar to the TVA value); right panel: a dust-rich disk with $\epsilon_0=1$.  
    \label{code_test_COS}}
\end{figure}

Fig. \ref{code_test_COS} shows the evolution of the maximum magnitude of gas velocity perturbations for the COS. For $\epsilon_0=0.01$ (left panel), there is a negligible difference between theoretical growth rates obtained from the TVA and exact treatment; we thus only plot the latter for clarity. The simulation growth rates, in either case, are in close agreement with theoretical expectations. 

\begin{figure}
    \centering
    \includegraphics[scale=0.455,clip=true, trim=0cm 0cm 0cm 0cm]{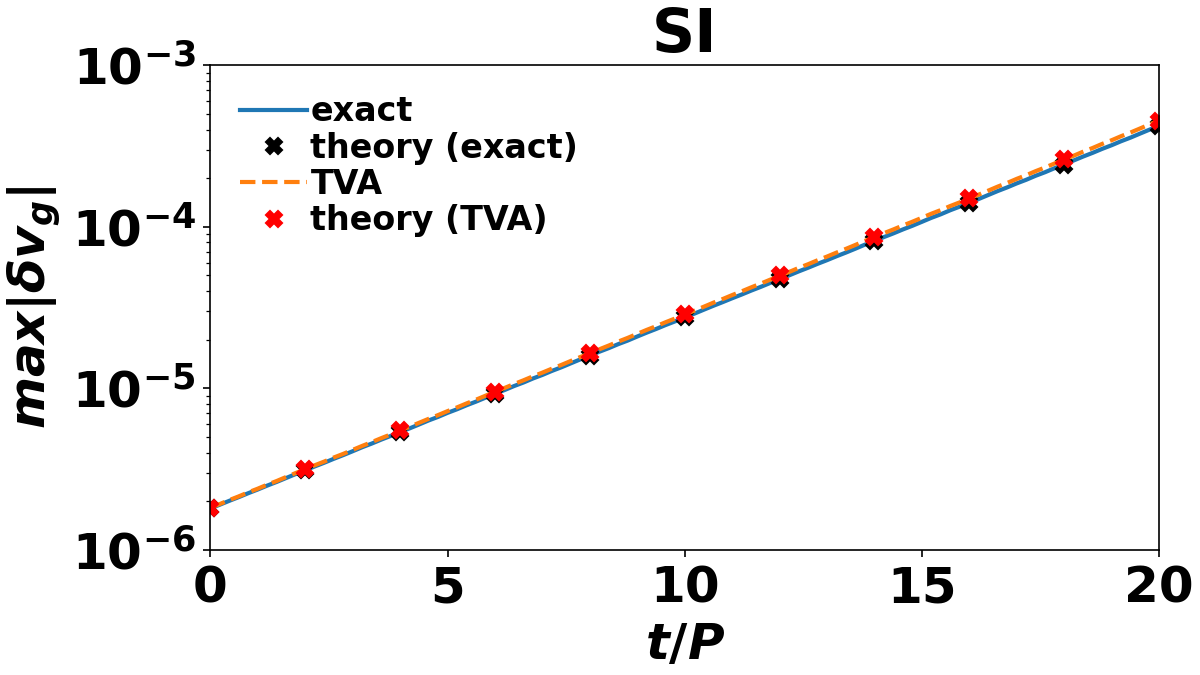}
    \caption{Linear growth of the SI simulated by \textsc{dedalus} (lines), compared with analytic growth rates (asterisks). Dust drag is treated via the full differential velocity equation (blue solid) or the TVA (orange dashed). 
    }
    \label{code_test_SI}
\end{figure}

The right panel Fig. \ref{code_test_COS} compares growth rates for the COS with $\epsilon_0=1$, which is weakened by dust feedback. Theoretical growth rates are again reproduced. However, the TVA underestimates growth rates by a factor of $\sim 2.4$. Fig. \ref{code_test_SI} shows the corresponding results for the SI. Here, TVA growth rates are marginally larger than the exact treatment, but simulation and theoretical values again agree. 

For completeness, Table \ref{code_test_table} compares the theoretical growth rates to that obtained from the \textsc{dedalus} runs. For the COS with $\epsilon_0=0.01$ and $\epsilon_0=1$, the simulation growth rates are measured between $t\in[0,50]\Porb$ and $t\in[0,300]\Porb$, respectively. For the SI run, we measure growth rates between $t\in[0,20]\Porb$. Our code implementation accurately reproduces theoretical growth rates with a maximum relative error of $O(10^{-4})$. 

\begin{deluxetable}{lcccc}
  \tablecaption{Selected growth rates of unstable modes. \label{code_test_table}}
    \tablehead{\colhead{Mode} & \multicolumn{2}{c}{Exact} & \multicolumn{2}{c}{TVA}\\
    \colhead{} &  \multicolumn{1}{c}{theory} & \multicolumn{1}{c}{simulation} & 
     \multicolumn{1}{c}{theory} & \multicolumn{1}{c}{simulation} 
    } 
\startdata
\hline\hline
COS ($\epsilon_0=0.01$) & 8.928487565399063e-03 & 8.928510894888546e-03 & 8.905393949389773e-03 & 8.905379670248154e-03 \\ 
COS ($\epsilon_0=1$) & 2.003292402239390e-03 & 2.003276214949109e-03 & 8.406434890964416e-04 & 8.406249376883315e-04 \\ 
SI & 4.333622586208401e-02 & 4.337942195298779e-02 & 4.391189786778842e-02 & 4.390035549322855e-02 
\enddata
\end{deluxetable}


\section{Resolution study}\label{resolution}
\begin{figure}
    \centering
    \includegraphics[scale=0.5]{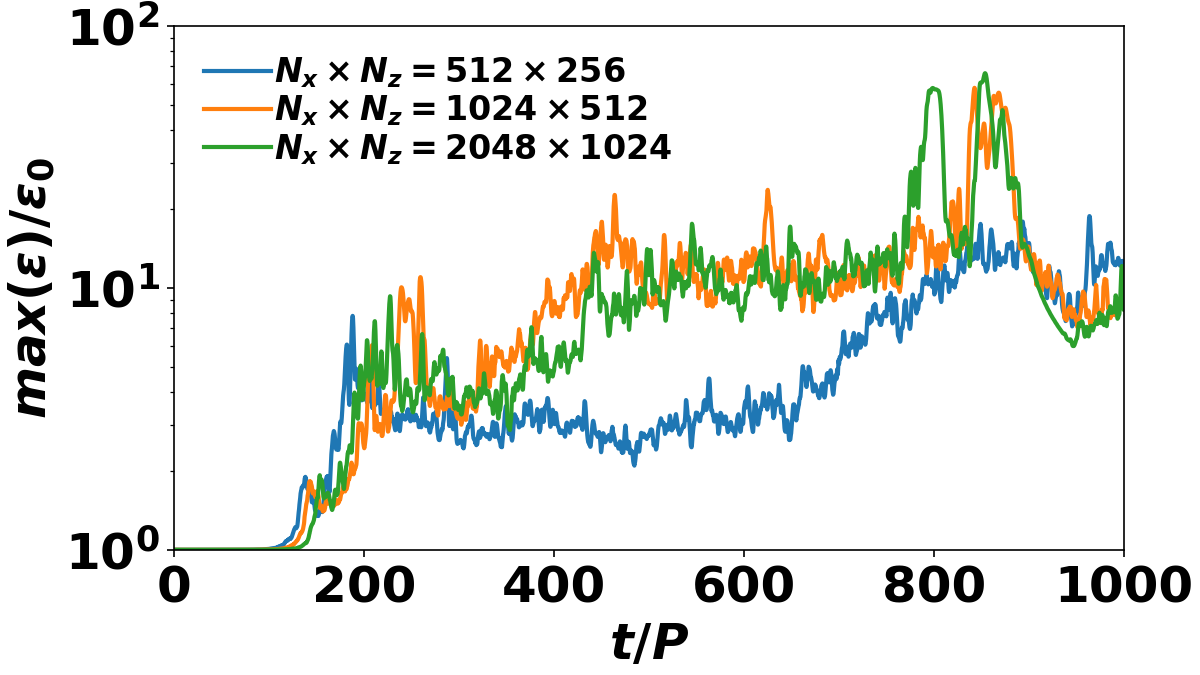}
    \caption{Maximum dust concentration factors at different resolutions, using the fiducial setup for the physical parameters (\S\ref{result_fid}). 
    }
    \label{resolution_study}
\end{figure}

Our fiducial resolution of $N_x\times N_z=2048\times 1024$ is limited by computational cost. Here, we perform lower-resolution runs to test for convergence. Fig. \ref{resolution_study} compares the fiducial run in \S\ref{result_fid} to that with half and a quarter of the resolution. Convergence is largely attained at $N_x\times N_z=1024\times 512$, while even lower resolutions yield notably reduced concentrations deep in the nonlinear regime ($t\gtrsim 300P$). 

\section{Dusty angular momentum flux in the linear regime}\label{linear_dusty_AMF}

We derive the AMF associated with linear dust velocity fluctuations in the limit of $\epsilon\to 0$, i.e., no feedback. That is, the flux induced by COS perturbations. We make use of the linearized horizontal gas momentum equations for the standard, inviscid COS:
\begin{align}
    &\sigma\dd \vgx = 2\Omega \dd \vgy - \left(\ii k_x\dd W + N_r^2\dd\theta\right), \label{COS_linear_vx} \\
    &\sigma\dd \vgy = -\frac{\Omega}{2}\dd\vgx, \label{COS_linear_vy}
\end{align}
which may be obtained from Eq. \ref{linear_vg} by neglecting drag terms, the background radial flow, and viscosity. These equations apply to axisymmetric, inviscid gas dynamics; only when we evaluate $\sigma$ do we specialize in the COS. It is convenient to write
\begin{align}
    \sigma = (s -\ii\omega)\Omega,
\end{align}
where $s$ and $\omega$ are real, dimensionless growth rates and frequencies, respectively.

The average gas AMF for linear perturbations of the form in Eq. \ref{linear_pert_form}, is given by $\Fg = \frac{1}{2}\re\left(\dd\vgx\dd \vgy^*\right)$, where $^*$ denotes the complex conjugate, and similarly for the dust AMF, $\Fd$. (These definitions differ from \citetalias{teed21} by a factor of four, which is immaterial to the discussion below.) Using Eq. \ref{COS_linear_vy}, we obtain
\begin{align}
    \Fg = -s \left|\dd\vgy\right|^2.
\end{align}
For unstable COS modes, $s>0$, and thus $\Fg$ is always negative, as discussed by \citetalias{teed21}.

We next calculate the average dust AMF, $\Fd = \frac{1}{2}\re\left(\dd\vdx\dd \vdy^*\right)$. We will make use of the TVA in the limit of zero feedback:
\begin{align}
    \dd\Delta v_x = \taus\left(\ii k_x\dd W + N_r^2\dd\theta\right),\label{Deltavx_pert_nofb}
\end{align}
to $O(\taus)$. This can be obtained from Eqs. \ref{TVA_expand} and \ref{Deltav0_pert} by setting $\epsilon\to 0$ and hence $\fgas\to 1$. Using $\dd\vdx = \dd \vgx + \dd \Delta v_x$ and the fact that $\dd \vdy = \dd \vgy$ in the TVA, we have 
\begin{align}
    \dd\vdx\dd\vdy^* &=  \dd\vgx\dd\vgy^* + \taus \dd\vgy^*\left(\ii k_x\dd W + N_r^2\dd\theta\right), \notag\\
    & = \dd\vgx\dd\vgy^* + \taus \dd\vgy^*\left(2\Omega \dd \vgy - \sigma\dd \vgx \right).
\end{align}
where we used Eq. \ref{Deltavx_pert_nofb} and \ref{COS_linear_vx} for the first and second equality, respectively. We then use Eq. \ref{COS_linear_vy} to eliminate $\dd\vgx$ and multiply by a half to obtain
\begin{align*}
     \frac{1}{2}\dd\vdx\dd\vdy^* = \frac{1}{2}\dd\vgx\dd\vgy^* + \st\left(1 + \frac{\sigma^2}{\Omega^2}\right)\left|\dd \vgy\right|^2.
\end{align*}
Finally, taking the real part gives
\begin{align}
\Fd &= \Fg + \st \left( 1 + s^2 - \omega^2\right)\left|\dd\vgy\right|^2\\
& = \left[\st \left( 1 + s^2 - \omega^2\right) - s \right]\left|\dd\vgy\right|^2.\notag
\end{align}
For nearly stationary modes with $s^2, \omega^2\ll 1$, we have $\Fd \simeq \Fg + \st |\dd\vgy|^2$, which is equivalent to Eq. \ref{Fd_model} in the main text\footnote{The factor of two difference stems from the fact that $\dd\vgy$ here is the complex amplitude of the perturbation, while that in Eq. \ref{Fd_model} is the complete perturbation.}.


The total AMF, $F = \Fg + \epsilon\Fd$, can be written as 
\begin{align}
    F = (1+\epsilon)\Fg + \epsilon\st \left( 1 + s^2 - \omega^2\right) \left|\dd\vgy\right|^2.
\end{align}
The COS corresponds to destabilized inertial waves, which have $|\omega|\leq 1$. Then, the second term is always positive, though smaller in magnitude compared to the first term because $\epsilon, \st\ll 1$. The total AMF is, therefore, still negative. 

\subsection{Low frequency limit}\label{linearAMF_low_freq}

For the COS, growth rates $s\ll 1$ since these are of $O(|N_r^2|/\Omega^2)$. Then in the low frequency limit with $|\omega|\ll 1 $, we have
\begin{align}
    \Fd =  \left(\st - s \right) \left|\dd\vgy\right|^2 \quad \text{(low frequency limit)}.
\end{align}
The dust AMF is thus positive (outward) if $\st > s$. This is the regime in our simulations with $s\lesssim O(10^{-2})$ but $\st = 0.1$. Notice also $\left|\Fd/\Fg\right| = \left|\st/s - 1\right|$, implying $|\Fd|\sim 10 |\Fg|$, as observed in simulations (Fig. \ref{compareAMF_epsilon}). 

Furthermore, if we equate the fluxes accounting for the dust-to-gas ratios, i.e., set $\left|\epsilon \Fd\right| = \left|\Fg\right|$, then the critical dust-to-gas ratio for the dust to affect the gas AMF is $\epsilon\sim 0.1$. This is roughly consistent with our simulations where zonal flows are suppressed at this $\epsilon$.

\subsection{High frequency limit}
On the other hand, for high-frequency modes $\omega^2\to 1$, so 
\begin{align}
    \Fd = -\left(1 - \st s \right)s \left|\dd\vgy\right|^2 \quad \text{(high frequency limit)},
\end{align}
which is generally negative because $\st , s \ll 1$. 

\bibliographystyle{aasjournal}
\bibliography{ref}

\end{document}